\documentclass[a4paper]{article}

\usepackage[margin=1in]{geometry}

\usepackage{amsfonts}
\usepackage{amsmath}
\usepackage{amssymb}
\usepackage{array}
\usepackage{bbm}
\usepackage{graphicx}
\usepackage[colorinlistoftodos]{todonotes}
\usepackage[colorlinks=true, allcolors=blue]{hyperref}
\usepackage[version=4, arrows=pgf-filled]{mhchem}
\usepackage{mathtools}
\usepackage{multirow}
\usepackage[semicolon]{natbib}
\usepackage{subcaption}
\usepackage{ulem}

\newcommand{\argmin}{\text{argmin}}

\newcommand{\vsubgrp}{\text{vsubgrp}}
\newcommand{\vgene}{\text{vgene}}
\newcommand{\naiveAA}{\text{naiveAA}}
\newcommand{\naiveAAclust}{\text{naiveAA-clust}}
\newcommand{\clust}{\text{clust}}
\newcommand{\neut}{\text{neut}}
\newcommand{\partis}{\texttt{partis}}

\newcolumntype{C}[1]{>{\centering\arraybackslash}m{#1}}

\newcommand{\beginsupplement}{%
        \setcounter{table}{0}
        \renewcommand{\thetable}{S\arabic{table}}%
        \setcounter{figure}{0}
        \renewcommand{\thefigure}{S\arabic{figure}}%
     }

\title{Predicting B Cell Receptor Substitution Profiles Using Public Repertoire Data}
\author{
  Amrit Dhar$^{1,2,*}$, Kristian Davidsen$^{2,*}$,  Frederick A. Matsen IV$^{2,\dag}$, Vladimir N. Minin$^{3,\dag}$\\ \\
  $^1$Department of Statistics, University of Washington, Seattle \\ 
  $^2$Fred Hutchinson Cancer Research Center\\
  $^3$Department of Statistics, University of California, Irvine \\
  $^{*}$joint first authors\\
  $^{\dag}$corresponding authors: \url{matsen@fredhutch.org}, \url{vminin@uci.edu}
}

\begin{document}
\maketitle

\begin{abstract}
B cells develop high affinity receptors during the course of affinity maturation, a cyclic process of mutation and selection.
At the end of affinity maturation, a number of cells sharing the same ancestor (i.e.\ in the same ``clonal family'') are released from the germinal center; their amino acid frequency profile reflects the allowed and disallowed substitutions at each position.
These clonal-family-specific frequency profiles, called ``substitution profiles'', are useful for studying the course of affinity maturation as well as for antibody engineering purposes.
However, most often only a single sequence is recovered from each clonal family in a sequencing experiment, making it impossible to construct a clonal-family-specific substitution profile.
Given the public release of many high-quality large B cell receptor datasets, one may ask whether it is possible to use such data in a prediction model for clonal-family-specific substitution profiles.
In this paper, we present the method ``Substitution Profiles Using Related Families'' (SPURF), a penalized tensor regression framework that integrates information from a rich assemblage of datasets to predict the clonal-family-specific substitution profile for any single input sequence.
Using this framework, we show that substitution profiles from similar clonal families can be leveraged together with simulated substitution profiles and germline gene sequence information to improve prediction.
We fit this model on a large public dataset and validate the robustness of our approach on an external dataset.
Furthermore, we provide a command-line tool in an open-source software package (\url{https://github.com/krdav/SPURF}) implementing these ideas and providing easy prediction using our pre-fit models.
\end{abstract}

\section*{Introduction}
In the therapeutic antibody discovery and engineering field, researchers commonly isolate antibodies from animal or human immunizations and screen for functional properties such as binding to a target protein.
Following the initial screening process, a small number of well-behaving antibodies (hits) are isolated for more rigorous examination of their biophysical properties in order to determine their potential as a therapeutic.
After this stage, only a few final antibodies remain as lead candidates.
However, even these carefully selected antibodies often have immunogenic peptides or other undesirable properties such as poor thermo/chemical stability and aggregation tendencies.
To address these problems, the art of antibody engineering has emerged \citep{igawa2011engineering}, with numerous rational design strategies developed to mitigate aggregation.
Researchers have removed hydrophobic surface patches to avoid aggregation \citep{clark2014remediating,casaz2014resolving,courtois2016rational,geoghegan2016mitigation}, ``deimmunized'' complementarity-determining regions by screening immunogenic peptides and mutating positions detrimental for peptide MHCII binding \citep{harding2010immunogenicity}, and improved thermostability through stable framework grafting \citep{mcconnell2014general} and targeted mutagenesis using predictions from proprietary structure/sequence analysis software \citep{seeliger2015boosting}.
Although referred to as ``rational'', the choice of which amino acid to use for a site-directed mutation is often made using 1) the germline as a reference, 2) biochemical similarity between amino acids, or 3) the highest probability amino acid from a generic substitution matrix (e.g.\ BLOSUM) \citep{henikoff1992amino}.
However, neither of these three methods are explicitly designed to conserve antibody functionality (i.e.\ binding to the same epitope with the same kinetics), so mutations are likely to have negative side effects on affinity.
These considerations motivate a prediction problem: given a B cell receptor (BCR) sequence, which positions can be modified, and to which amino acids, without drastically changing the binding properties of the resulting BCR?

An immunization-derived antibody has already implicitly explored the mutational space through the population of B cells from which it derives, referred to as its clonal family (CF).
The members of a CF are raised during affinity maturation in a germinal center and carry fitness information about the effect of amino acid substitutions.
A profile of the observed substitutions aggregated over all the B cells in a CF reveals which sites are more conserved, which sites can be more freely edited, and which amino acids can be used for replacements.
However, we generally do not sequence all the B cells that are released from a germinal center so the information to make such a substitution profile is lost.
Thus, we can formulate a more specific version of our prediction problem: given bulk BCR data and a single input sequence, can we infer the most likely per-site substitutions that are allowed in its true germinal center lineage?

We begin by reviewing the natural mutation and selection process of germinal center affinity maturation.
The Darwinian selection undertaken inside a germinal center is driven by B cells' ability to bind the antigen through the membrane-embedded BCR.
The population of B cells in a germinal center is under stringent selection while being highly mutated, driving the cell population towards higher and higher affinity until the germinal center is dissolved.
Each germinal center is seeded by around one hundred naive B cells, but eventually internal competition makes one or a few of these lineages take over the whole germinal center \citep{tas2016visualizing}.
Although B cells in the germinal center reaction experience an extraordinarily high mutation rate ($10^6$ fold higher than the regular somatic mutation rate \citep{victora2012germinal}), they rarely harbor more than 15\% mutations at the DNA level \citep{Briggs2017}.
However, since they must maintain some degree of antigen specificity to survive during the course of the germinal center reaction, lineages evolve in small incremental steps \citep{kepler2014reconstructing,kuraoka2016complex} and therefore, even lineages that drift far away from their naive B cell ancestor most likely maintain the same epitope specificity throughout the germinal center reaction \citep{Schmidt2013-jw}.

We can describe the combination of germinal center mutation and selection dynamics by computing per-site amino acid frequency vectors from observed BCR sequence data.
We follow previous authors in calling site-specific amino acid probability vectors ``substitution profiles'', where each vector in a profile stores the probabilities of observing the 20 different amino acids at a given site \citep{sheng2017gene}.
We use the concept of a clonal family, defined by a shared naive DNA sequence, to segment BCR sequences into evolutionarily-related groups \citep{ralph2016likelihood}.
CF inference is highly informed by nucleotide sequences and therefore performed using DNA sequences.
This makes DNA-level information necessary even though germinal center selection operates at the protein level and synonymous codons do not possess any fitness advantages (modulo transcription rate differences and codon bias, which we follow many others in ignoring here).
The per-site amino acid frequency vectors described above form the substitution profile estimates; the substitution profile estimates converge to the true substitution profiles as the number of sequences sampled from the same CF tends to infinity.

\begin{figure}[ht!]
\centering
\includegraphics[width=\textwidth]{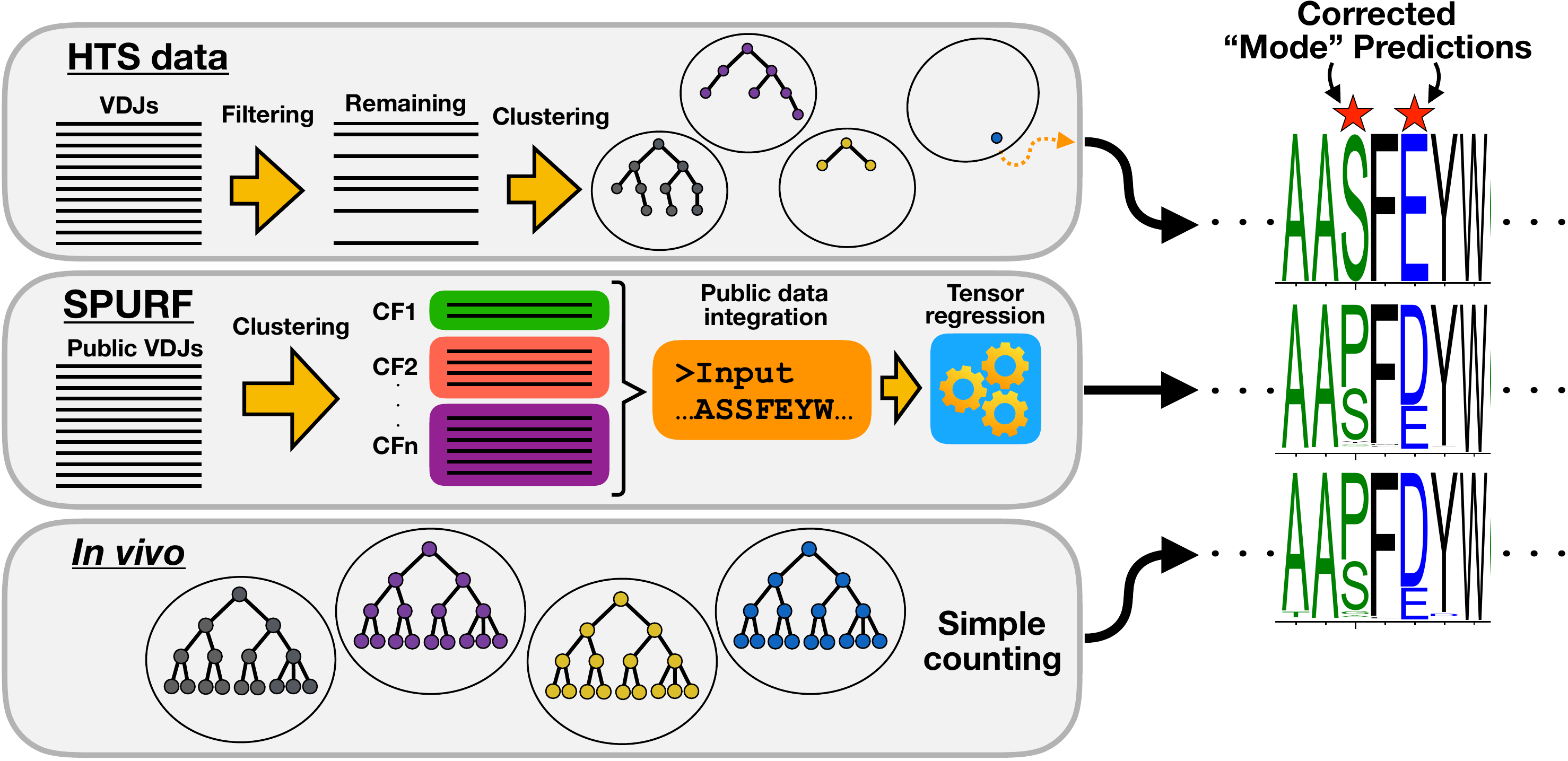}
\caption{
Amino acid substitution profiles viewed from three different perspectives:
High-throughput sequencing data (\underline{HTS data}) yields large amounts of VDJ sequences, but because of uneven sampling many CFs will be sampled just once, resulting in poor representations of the amino acid substitution profiles of those true CFs.
``Substitution Profiles Using Related Families'' (\underline{SPURF}) is a statistical framework that integrates large scale Rep-Seq data to predict amino acid substitution profiles for singleton CFs.
\underline{\textit{In vivo}} affinity maturation will test many different mutations and the resulting CFs reflect the amino acid substitution profiles that we attempt to predict.
}
\label{fig:overview_figure}
\end{figure}

Most CFs do not contain enough sequences in order to get a detailed substitution profile estimate.
Indeed, most CFs in repertoire sequencing (Rep-Seq) samples have few members and a large fraction are singletons due to the exponential nature of the CF size distribution \citep{ralph2016likelihood}.
Additionally, many antibody screening methods are not geared towards whole repertoire sequencing.
One may wish, then, to enhance the substitution profile estimates for data-sparse CFs with substitution profile information from similar CFs.

In this paper, we present ``Substitution Profiles Using Related Families'' (SPURF), a penalized tensor regression framework that integrates multiple sources of information to predict the CF-specific amino acid frequency profile for a single input BCR sequence (\autoref*{fig:overview_figure}).
Some of these information sources include substitution profiles for CFs in large, publicly available BCR sequence datasets and germline gene sequence information.
We combine the local context-specific profile information with global profile information derived from other related germinal centers by regularizing the noisy local substitution profile estimate and pooling it closer towards more robust global profile estimates.
Even though each germinal center focuses on binding to a unique epitope context, there are structural and possibly functional properties associated with BCR sequences that are common across germinal centers that we can leverage.

In addition, our inference machinery uses both standard and spatial lasso penalties as model regularizers and, as a result, furnishes sparse, interpretable parameter estimates.
While our output type shares some similarities to that described by \citet{sheng2017gene}, the proposed objective, approach, and details differ (e.g.\ they predict substitution profiles for gene families, we predict substitution profiles for CFs).
We enable substitution profile prediction for single input BCR sequences based on profiles derived from a high-quality repertoire dataset that contains B cell samples from many human donors.
To demonstrate the usefulness of our technique, we validate SPURF on an external dataset containing CFs extracted from a single human donor.
Lastly, we implement SPURF in an open-source software package (\url{https://github.com/krdav/SPURF}), which outputs a predicted CF-specific substitution profile and an associated logo plot based on a single input BCR sequence.

\section*{Methods}

\subsection*{Overview}
The aim of our model is to take a single sequence and predict the substitution profile for the CF from which this single sequence derived.
For this prediction problem, we have no direct information about this desired substitution profile other than the information contained in the input sequence itself, but we may use other information (e.g.\ from the inferred germline gene, simulated substitutions, or information derived from published BCR sequence datasets).
For large CFs, a CF-specific substitution profile can be constructed simply by counting and making a per-site frequency matrix, with the rows of the matrix representing each of the 20 amino acids, and the columns being the sequence positions.

For training, we extract a collection of such large CFs and use them to build ``ground truth'' CF-specific substitution profiles as a training set for fitting the model.
A randomly sampled single sequence is then taken out from each of these large CFs to predict the substitution profile, which is compared to the ground truth.
We refer to these single sequences, sampled from large CFs, as subsamples.

\begin{figure}[ht!]
\centering
\includegraphics[width=1.0\textwidth]{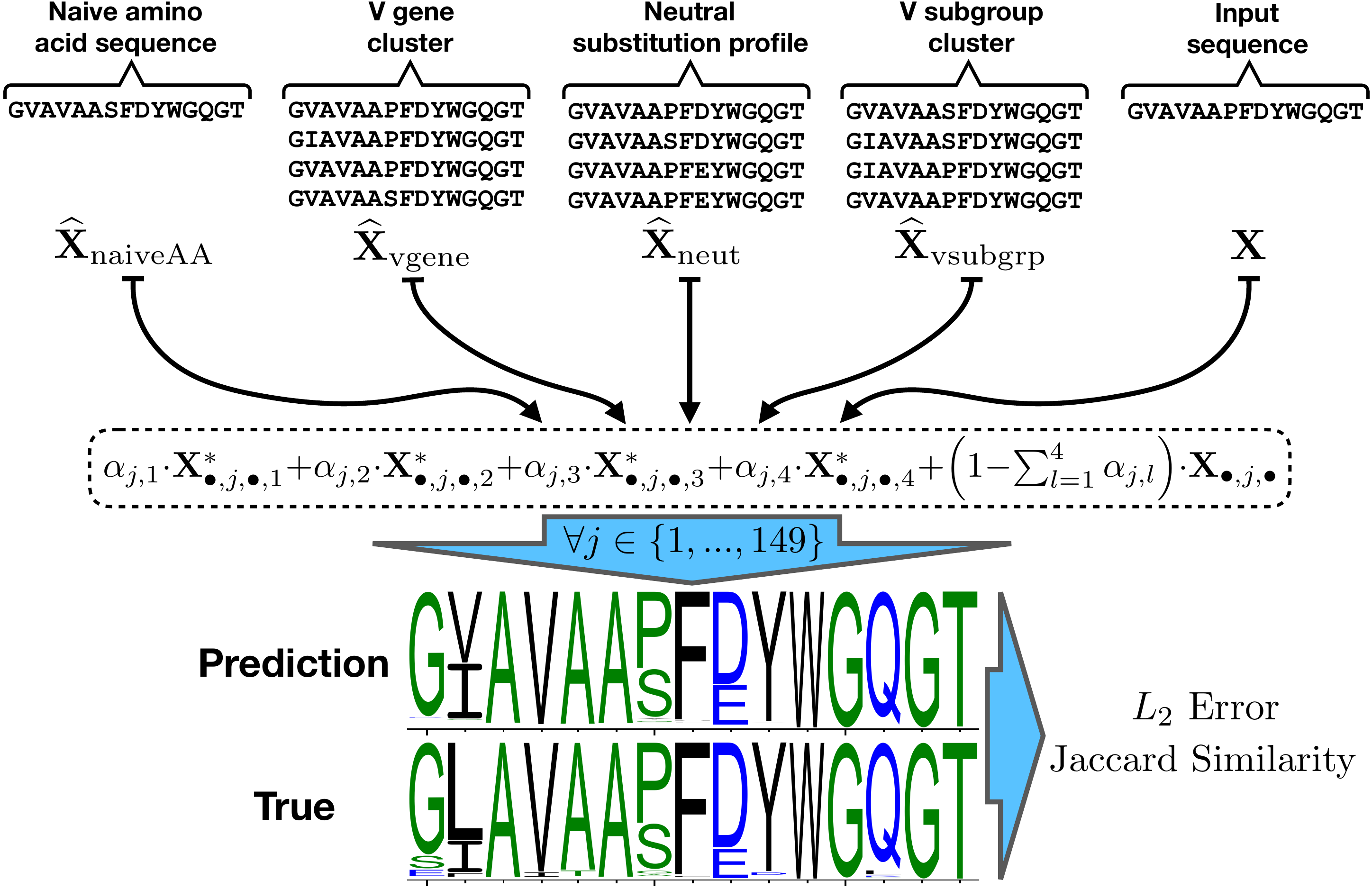}
\caption{
SPURF uses a per-site linear combination of substitution profiles from diverse sources to predict complete substitution profiles from a single member of a CF.
At the top are the different profiles that serve as inputs to the model, some directly related to the naive sequence ($\widehat{\mathbf{X}}_{\naiveAA}$ and $\widehat{\mathbf{X}}_{\neut}$), and others partitions of the public Rep-Seq datasets ($\widehat{\mathbf{X}}_{\vgene}$ and $\widehat{\mathbf{X}}_{\vsubgrp}$).
To predict a substitution profile, a weighted average is taken over the input sequence $\mathbf{X}$ and external profiles $\mathbf{X}^* = \bigl\{\widehat{\mathbf{X}}_{\naiveAA}, \widehat{\mathbf{X}}_{\vgene}, \widehat{\mathbf{X}}_{\neut}, \widehat{\mathbf{X}}_{\vsubgrp}\bigr\}$ (see the dashed line bubble).
The vertical blue arrow indicates that the weighted average (in the dashed line bubble) occurs at each of the 149 AHo positions.
Once a predicted profile is generated, this is compared to ground truth using either $L_2$ error or Jaccard similarity as a performance metric.
The $\boldsymbol{\alpha}$ vectors are estimated by optimizing the objective function, which also includes a statistical regularization term to prevent overfitting (not shown for simplicity).
}
\label{fig:model_overview}
\end{figure}

To make the best possible prediction, we need a flexible model framework that can accommodate different sources of information seamlessly (\autoref*{fig:model_overview}).
For example, previous work by \citet{sheng2016effects} and \citet{Kirik2017-bc} suggests that the various V genes have different characteristic paths of diversification.
We can obtain a data-driven summary of that intuition by building profiles from large Rep-Seq data sets stratified by V gene.
We may also think that the neutral substitution process is an important factor in determining substitution profiles \citep{sheng2016effects}.
We can quantify that sort of information by repeatedly simulating the neutral substitution process using a context-sensitive model \citep{cui2016model}.

To make predictions using these types of information, we need a way of describing the various sites, and a way of integrating the information across the sites.
We use the AHo numbering scheme \citep{honegger2001yet} to provide a single coordinate system to all sequences via its fixed-length numbering vector going from 1 to 149.
Given this coordinate system, we use a site-wise weighted average of the input predictive profiles using a $\boldsymbol{\alpha}$ weight vector for each source of profile information.

To train this model, we fit the $\boldsymbol{\alpha}$ vectors by minimizing some objective function that quantifies the difference between the predicted profiles (where the prediction uses the subsampled sequence and the external profile information) and the ``ground truth'' substitution profiles from the large CFs.
Any objective function could be used, but here we provide implementations of two such functions, a ``fine-grained'' $L_2$-error-based objective and a ``coarse-grained'' Jaccard-similarity-based objective \citep{jaccard1912distribution}.

We use two forms of regularization to avoid overfitting the many parameters of this model.
This includes a standard lasso penalty to shrink weights to zero that do not contribute significantly to prediction performance \citep{tibshirani1996regression}.
We also use a fused lasso penalty \citep{tibshirani2005sparsity, tibshirani2014adaptive} to smooth differences between parameters at nearby sites in the sequence.
These regularization terms have tuning parameters that regulate the strength of the penalties and are estimated using cross-validation.

Given this setup, a forward stepwise selection procedure is run with cross-validation to pick the set of external profiles to use in the final model.
As a last check, this model is tested on an external dataset to give a fair estimate of the prediction performance.

\subsection*{Data}
We divide input data into two parts, with each part for a respective purpose: 1) model fitting and model testing and 2) providing ``public'' substitution profiles over clustered data to be used by our model.
Throughout this work, we are careful to not use the same data for both purposes as this would bias our estimates; as a final validation, we test SPURF on an external dataset which is only used in this validation.
Because we do not model sequence error, we only include high-quality data that we have high confidence in.
We collect post-processed data files from 6 published works on Rep-Seq, which we refer to as repertoire data 1 to 6 (RD1-6):
\begin{enumerate}
  \item RD1 from \citet{gupta2017hierarchical}, which is an Illumina MiSeq re-sequencing of the samples in \citet{laserson2014high}, where they sequence multiple time-points before and after influenza vaccination of 3 donors using the 454 pyrosequencing platform.
  \item RD2 from \citet{vander2017dysregulation}, from a study of the auto-immune disease Myasthenia Gravis (MG), in which 9 MG patients and 4 healthy donors participate.
  \item RD3 from \citet{stern2014b}, containing data from different tissues in a study of B cell response in 4 multiple sclerosis patients.
  \item RD4 from \citet{tsioris2015neutralizing}, from a study of neutralizing antibodies against the West Nile virus by sequencing naive and memory cells from 7 virus infected donors.
  \item RD5 from \citet{shugay2014towards}, from a study of Rep-Seq error correction by sequencing naive, plasma, and memory cells from a single healthy donor.
  \item RD6 from \citet{meng2017atlas}, from the ``B cell tissue atlas'' acquired from the ImmuneDB web portal.
\end{enumerate}
All datasets are acquired in their post-processed form with read processing performed as described in their respective publications.

The first five datasets (RD1-5) are prepared from unique molecular identifier (UMI) barcoded cDNA spanning the whole VDJ region and sequenced on the Illumina MiSeq platform using overlapping paired-end reads.
Using the UMI, these reads are processed to address both PCR and sequencing errors giving high confidence reads \citep{shugay2014towards}.
Briefly, UMIs are used for error correction in conjunction with either of the pRESTO \citep{vander2014presto} or MIGEC \citep{shugay2014towards} processing pipelines and an appropriate Phred quality score cutoff.
Paired-end reads are assembled using pRESTO and only the set of high confidence assembled reads constitute the final dataset used in this work.
RD6 is the only dataset not prepared with UMIs; however, it is sequenced directly from genomic DNA (gDNA) instead of the more common practice of sequencing mRNA.
Sequencing gDNA has the benefit of avoiding mutations introduced by the transcription machinery as well as mutations introduced in the RT-PCR step.
On the other hand, DNA sequencing is not able to discriminate between expressed versus unexpressed BCRs (e.g.\ in the case of faulty VDJ recombination) and therefore we apply aggressive filtering of non-functional BCR sequences.
We prefer quality over quantity and therefore avoid datasets from the 454 technology because of their higher indel frequencies compared to those from Illumina technologies \citep{loman2012performance}.

Individual sequence files are merged based on donor identity so that the number of sample files matches the number of donors; this process yields 33 donor files.
The donor files are then annotated and partitioned into CFs using the \partis\ software \citep{ralph2016consistency,ralph2016likelihood}.
Each donor file is run separately from the other files so CFs are defined by their unique \partis-inferred naive sequence and donor identity.
To ensure we obtain the highest quality and most biologically relevant sequences, \partis\ is run in its most restrictive mode, discarding all reads with VDJ recombinations that are deemed as unproductive because of out-of-frame N/P junction nucleotides, missing invariant codons, or stop codons inside the VDJ region; furthermore, the most accurate \partis\ partitioning mode (``full'') is used to get the best CF estimates.
Lastly, productive VDJ-recombined sequences are removed if they contain indels to assure concordance between the length of the naive sequence and the length of the read sequences in its CF.

At this stage, some sequences contain ambiguous bases (e.g., because of primer masking); these are allowed to pass only if the ambiguous bases are inside the first or last 30 nucleotides of the VDJ region (equivalent to the length of the potentially masked PCR primers), otherwise they are discarded.
This is a way of substituting the error-prone ends with neutral bases that minimize variance and maintain a conservative estimate of the substitutions; we also note that this has no apparent effect on the subsequently-described estimates (\autoref*{fig:alpha_profile_plot_collapsed} and \autoref*{fig:pred_fwkcdr}).
For all sequences that pass this requirement, ambiguous bases are substituted with bases from the naive sequence in batches of 3 nucleotides (i.e.\ one codon) at a time until all ambiguous bases are resolved.
Sequences are then translated into their respective amino acid sequences and de-duplication of repeated amino acid sequences is done within each CF.
Because our statistical methodology operates on these amino acid sequences, we use the word ``sequence'' in subsequent sections to refer to these amino acid sequences.
All CFs with fewer than 5 unique sequences are discarded.
From these remaining CFs, their inferred naive sequences are used for antibody sequence numbering with the ANARCI software \citep{dunbar2015anarci} under the AHo numbering scheme \citep{honegger2001yet}.
As a result of our restriction to non-indel sequences, all sequences within a given CF have equal length; thus, the AHo numbering from the naive sequence can be positionally transferred to all its CF-related read sequences.
Finally, for each CF, the amino acid usage is extracted as a vector of counts at each AHo position.
This overall dataset, which we call the ``aggregated'' dataset, contains 518,174 sequences distributed over 31,893 CFs and is built as a matrix of counts with rows denoting CFs and columns representing AHo positions and amino acid identities.
All data used to build this aggregated dataset is public and freely available.
We provide processed data partitioned into CFs upon request.

\subsubsection*{Model Fitting Dataset}
To fit our CF-specific substitution profile prediction model, it is desirable to use the CFs from the aggregated dataset with the most sequence members so we can train using the observed substitution profiles with the least amount of noise; on the other hand, it is also desirable to extract CFs from as many donors as possible to avoid overfitting towards a few similar donors.
To achieve both goals, we pick 500 CFs as a ``model fitting'' dataset as follows.
We first exclude any CFs with less than 100 sequences from being eligible to be picked.
We then cycle through donors, each time picking the largest remaining eligible CF.
If a donor does not have any remaining eligible CFs, it is skipped.
The process ends when 500 CFs are found; all unpicked CFs are used as the ``public'' dataset.

In addition, we perform subsampling for each CF in the model fitting dataset; this is the information from which we would like to predict the full profile.
First, a single sequence is randomly chosen from each CF, then \partis\ is re-run using each of these subsampled sequences to re-do the VDJ annotation and naive sequence inference.
For some inferred naive DNA sequences, a stop codon is incidentally present in the N/P nucleotides of the junction region; these are considered spurious and replaced by the identically positioned codon from the input sequence.
We stress that the CF-specific annotation and naive sequence are inferred solely based on the subsampled sequence itself and are not determined using information from the other CF sequence members.
Additionally, the parameters used within the \partis\ clustering and annotation procedure are derived from an external dataset.
Once we finish the \partis\ inference process on the subsampled sequences, we construct the amino acid count matrix for these same sequences; we denote these substitution profiles as the ``subsampled'' profiles because they are subsampled from the ``full'' profiles in the model fitting dataset.

\subsubsection*{Simulation of Neutral Substitution Profiles}
For each of the 500 subsampled substitution profiles, we also simulate a neutral substitution profile via a context-sensitive model.
For each subsampled sequence, we calculate its number of somatic hypermutations (SHMs) and introduce that number of mutations sequentially into the inferred naive DNA sequence according to the BCR-specific neutral substitution model S5F \citep{cui2016model}.
Once the last mutation is introduced, the simulated DNA sequence is translated into an amino acid sequence and stored as a sample of the neutral substitution process.
This procedure is repeated 10,000 times and the count profile aggregated over all the samples is referred to as the ``neutral'' profile.

\subsubsection*{External Validation Dataset}
For validation, a test set, called ``Briggs'', is made from the healthy donor single cell droplet sequencing dataset described in \citep{Briggs2017}.
Briefly, the data is made by passing 3 million B cells into 6 emulsion pools, each droplet with a unique barcode, and then reverse transcribing mRNA inside these droplets, attaching both a droplet and a molecular barcode.
After breaking the emulsion, cDNA is sequenced and processed using UMI consensus building using pRESTO.
The highest-quality UMI consensus sequence is extracted from each drop and aggregated into the final heavy chain dataset, which is then further partitioned into CFs using \partis.
Finally, the validation dataset is built up in the same manner as the model fitting dataset, where the only difference is that we allow smaller CFs to enter this dataset (minimum 28 sequences) in order to increase the number of extracted CFs to 100.
For this external dataset, sequences and processed data partitioned into CFs are available upon request.
\begin{table}[ht!]
\centering
\begin{tabular}{lcccccc}
        & \multicolumn{6}{c}{Dataset summary statistics}      \\ \cline{2-7}
Dataset       & $N_{\text{donors}}$  & $N_{\text{CF}}$  & Total $N_{\text{seq}}$  & Min $N_{\text{seq}}$ & Median $N_{\text{seq}}$ & Max $N_{\text{seq}}$ \\ \hline
Aggregated    & 33        & 31,893 & 518,174       & 5          & 9             & 2,709      \\
Model fitting & 15        & 500    & 98,887        & 100        & 147           & 2,709      \\
Public        & 33        & 31,393 & 419,287       & 5          & 8             & 104        \\
Briggs        & 1         & 100    & 6,702         & 28         & 44            & 370        \\ \hline
\end{tabular}
\caption{
Number of donors ($N_{\text{donors}}$), number of CFs ($N_{\text{CF}}$), number of sequences from all CFs (Total $N_{\text{seq}}$), smallest CF size (Min $N_{\text{seq}}$), median CF size (Median $N_{\text{seq}}$), and maximum CF size (Max $N_{\text{seq}}$).
``Aggregated'' is the base dataset aggregating RD1-6.
``Model fitting'' refers to the dataset with the 500 largest CFs from the ``Aggregated'' dataset.
``Public'' is the dataset left after the ``Model fitting'' dataset is extracted from the ``Aggregated'' dataset.
``Briggs'' is the external validation dataset used for testing.
}
\label{table:ds_sumstats}
\end{table}

\subsection*{Input Data Tensor}
Before we present our penalized tensor regression model, we first describe how the input data for the model is constructed, building off the data descriptions in the last subsection.
Throughout the rest of this section, we assume the count matrices are normalized to frequencies and reorganized into three-dimensional tensors (i.e.\ arrays) as follows.
For any substitution profile tensor $\mathbf{T} = \{T_{i,j,k}\}$, let $T_{i,j,k}$ denote the substitution frequency of the $k$th amino acid at the $j$th AHo position for the $i$th CF; we represent the subsampled, full, and public substitution profile tensors as $\mathbf{X}$, $\mathbf{Y}$, and $\mathbf{Z}$, respectively.
Our goal is to use the subsampled profiles $\mathbf{X}$ to predict the corresponding full substitution profiles $\mathbf{Y}$ (i.e.\ we want to construct a function $F(\mathbf{X})$ such that $F(\mathbf{X}) \approx \mathbf{Y}$).
We incorporate information from the public dataset $\mathbf{Z}$ to enhance these predictions.
In addition to the subsampled profiles, we use other types of substitution profiles within $F(\mathbf{X})$:
\begin{enumerate}
\item Public substitution profiles segmented by the inferred V-subgroup label ($\widehat{\mathbf{X}}_{\vsubgrp}$);
\item Public substitution profiles segmented by the inferred V-gene label ($\widehat{\mathbf{X}}_{\vgene}$);
\item Inferred naive sequence ``substitution profiles'' ($\widehat{\mathbf{X}}_{\naiveAA}$);
\item Public substitution profiles segmented by the inferred naive sequence ($\widehat{\mathbf{X}}_{\naiveAAclust}$);
\item Public substitution profiles segmented by the original frequency profiles ($\widehat{\mathbf{X}}_{\clust}$);
\item Neutral substitution profiles ($\widehat{\mathbf{X}}_{\neut}$).
\end{enumerate}
To compute the external profiles in $\widehat{\mathbf{X}}_{\vsubgrp}$ (resp.\ $\widehat{\mathbf{X}}_{\vgene}$), we cluster the public dataset $\mathbf{Z}$ by averaging its CF-specific substitution profiles according to the \partis-inferred \citep{ralph2016consistency} IMGT defined \citep{lefranc2001nomenclature} V-subgroup (resp.\ V-gene) labels and then assign each row in $\mathbf{X}$ to a V-subgroup (resp.\ V-gene) cluster profile according to its V-subgroup (resp.\ V-gene) identity.
We obtain the second set of profiles $\widehat{\mathbf{X}}_{\naiveAA}$ by using the \partis-inferred naive sequences as substitution profiles (these profiles contain zeros and ones because they are based on one sequence only); we re-emphasize that these naive sequences are inferred based only on the corresponding subsampled sequences in $\mathbf{X}$.
We cluster the public dataset $\mathbf{Z}$ once more by running K-means clustering based on the inferred naive sequences in $\mathbf{Z}$ and obtain our third set of substitution profiles $\widehat{\mathbf{X}}_{\naiveAAclust}$ by assigning each CF in $\mathbf{X}$ to its closest cluster centroid.
The additional cluster profiles $\widehat{\mathbf{X}}_{\clust}$ are obtained similarly as above, except in this case, we run K-means clustering based on the original frequency profiles in $\mathbf{Z}$.
The K-means clustering procedure is run over a grid of cluster sizes ranging from 2 to 120 using the algorithm described by \citet{hartigan1979algorithm} with the standard euclidean distance metric.
Lastly, the tensor $\widehat{\mathbf{X}}_{\neut}$ contains the simulated S5F neutral substitution profiles, which are described in the previous subsection.

The frequency tensors $\widehat{\mathbf{X}}_{\vsubgrp}$ and $\widehat{\mathbf{X}}_{\vgene}$ are important to include in our analysis because these profiles capture substitution information at the level of the V subgroup (V1, V2, ...) and V gene (V1-5, V2-2, ...), respectively; this is similar to the types of profiles obtained in \citep{sheng2017gene}.
As described in the introduction, most germinal center lineages do not accumulate many mutations relative to the naive sequence so substitution profiles based solely on the naive sequence (like $\widehat{\mathbf{X}}_{\naiveAA}$) may be informative for predicting the mutational patterns at conserved residue positions.
In addition, we believe that the $\widehat{\mathbf{X}}_{\naiveAAclust}$ cluster profiles are useful as the naive sequence can greatly influence the pattern of substitutions in a CF due to local sequence context.
Unlike the $\widehat{\mathbf{X}}_{\vsubgrp}$ and $\widehat{\mathbf{X}}_{\vgene}$ substitution profiles, which are based on IMGT labeling schemes, the profiles in $\widehat{\mathbf{X}}_{\naiveAAclust}$ (and $\widehat{\mathbf{X}}_{\clust}$) are determined by a data-driven clustering procedure, which allows us to group CFs in $\mathbf{Z}$ in a more intricate fashion.
The simulated neutral substitution profiles $\widehat{\mathbf{X}}_{\neut}$ are able to provide some insight into the CF-specific SHM processes without the corresponding clonal selection effects.

To condense our model presentation, we introduce a four-dimensional tensor $\mathbf{X}^*$ that combines as many of the input profiles mentioned previously as we would like, where $p$, the size of the fourth tensor dimension, represents the number of external profiles used.
We define $\mathbf{X}^* \equiv \{X^*_{i,j,k,l} \}$ to be the input data tensor that incorporates all the external information we want to use in our substitution profile predictions; note that $i \in \{ 1, ..., N_{CF} \}$ ($N_{CF}$ CFs in the tensors), $j \in \{ 1, ..., 149 \}$ (149 AHo positions), $k \in \{ 1, ..., 20 \}$ (20 amino acids), and $l \in \{ 1, ..., p \}$ ($p$ external profiles).
Each element $X^*_{i,j,k,l}$ represents a substitution frequency as described above for $\mathbf{T}_{i,j,k}$; for instance, $X^*_{5,130,1,4}$ represents the substitution frequency of the first amino acid (i.e.\ alanine) at the 130th AHo position for the 5th CF in the 4th profile in the tensor.
In addition, we use the indexing symbol $\bullet$ to extract all elements of a particular array dimension of a tensor (i.e.\ $\mathbf{X}^*_{10, 50, \bullet, 2}$ specifies the full substitution profile of the 20 amino acids at the 50th AHo position for the 10th CF in the 2nd profile in the tensor).
This setup allows us to easily include as many external profiles as we would like.

\subsection*{Model Formulation}
Given the subsampled profiles $\mathbf{X}$ and all the external profiles $\mathbf{X}^*$, we compute a weighted average to form an estimator of $\mathbf{Y}$.
Our independent-across-sites model $F(\mathbf{X}) = \bigl[ f(\mathbf{X}_{\bullet,1,\bullet}), ..., f(\mathbf{X}_{\bullet,149,\bullet}) \bigr]$ is specified as follows:
\begin{equation}
f(\mathbf{X}_{\bullet,j,\bullet}) \equiv f(\mathbf{X}_{\bullet,j,\bullet}; \boldsymbol{\alpha}_{j,\bullet}) = \sum_{l=1}^p \alpha_{j,l} \cdot \mathbf{X}^*_{\bullet,j,\bullet,l} + \Bigl( 1 - \sum_{l=1}^p \alpha_{j,l} \Bigr) \cdot \mathbf{X}_{\bullet,j,\bullet},
\end{equation}
where $\boldsymbol{\alpha} = \{ \alpha_{j,l} \}$; $0 \leq \alpha_{j,l} \leq 1$; $0 \leq \sum_{l=1}^p \alpha_{j,l} \leq 1$ represents the site-specific weights of the different external profiles for $j = 1, ..., 149$ and $l = 1, ..., p$.
Although we consider $f$ to be a function of the per-site data $\mathbf{X}_{\bullet,j,\bullet}$, the frequencies $\mathbf{X}^*_{\bullet,j,\bullet,l}$ are computed using sequence-level, site-dependent information.
With $149 \times p$ parameter values of $\boldsymbol{\alpha}$, this is a highly parameterized model so we include regularization terms to prevent overfitting and obtain sparse, interpretable parameter estimates.
Specifically, we use standard and spatial (fused) lasso penalties to achieve these goals.

Standard lasso penalties shrink individual parameters to zero and are commonly used to obtain sparse solutions in regression problems \citep{tibshirani1996regression}.
It has been shown that regression models using standard lasso penalties provide more accurate predictions than models using best subset selection penalties when there is a low signal-to-noise ratio \citep{hastie2017extended}, which probably holds true in our problem as well.
In addition, standard lasso penalties are convex functions, which is important in a regression problem as it guarantees that a local minimum is indeed a unique global solution \citep{boyd2004convex}.

On the other hand, fused lasso penalties shrink the differences between parameters to zero and are useful in regression problems with spatially-related covariates \citep{tibshirani2005sparsity}.
We believe that the $\boldsymbol{\alpha}$ parameters have a spatial relationship (i.e.\ adjacent residues are under similar constraints); for instance, given that the mutations in the framework regions are largely related to antibody stability, it makes sense that we would weight external profile information similarly in those regions.
The fusion penalty in this setting enforces smoothness of the $\boldsymbol{\alpha}$ trend across the AHo positions.
For example, if we penalize first-order differences of the $\boldsymbol{\alpha}$ trend, the fitting procedure will necessarily favor trends that have no slope (i.e.\ that are piecewise constant).
We can obtain more flexible piecewise polynomial $\boldsymbol{\alpha}$ trends by penalizing higher-order successive differences of $\boldsymbol{\alpha}$ \citep{tibshirani2014adaptive}.

In our modeling framework, the standard lasso penalty is represented as $\sum_{j=1}^{149} \sum_{l=1}^p |\alpha_{j,l}| = \bigl|\bigl| \boldsymbol{\alpha} \bigr|\bigr|_1$ and the fused lasso penalty is specified by $\sum_{l=1}^p \bigl|\bigl| \nabla^d (\boldsymbol{\alpha}_{\bullet,l}) \bigr|\bigr|_1$, where $|| \cdot ||_q$ denotes the $L_q$ norm and $\nabla^d ( \cdot )$ represents the $d$th difference operator.
This $\nabla^d ( \cdot )$ operator accepts a vector $\mathbf{v}$ as input (call its length $n_{\mathbf{v}}$) and outputs a length-$(n_{\mathbf{v}} - d)$ vector that results from successively differencing adjacent elements $d$ times.
In the special case when $d = 1$, the fusion penalty becomes $\sum_{l=1}^p \bigl|\bigl| \nabla^1 (\boldsymbol{\alpha}_{\bullet,l}) \bigr|\bigr|_1 = \sum_{j=2}^{149} \sum_{l=1}^p |\alpha_{j,l} - \alpha_{j-1,l}|$; the  $|\alpha_{j,l} - \alpha_{j-1,l}|$ terms can be interpreted as first-order discrete derivatives.

Our unpenalized objective function can be written as:
\begin{equation}
L_2^{\boldsymbol{\alpha}} \equiv L_2^{\boldsymbol{\alpha}}(\mathbf{Y}, F(\mathbf{X})) = \frac{1}{149 \cdot N_{CF}} \sum_{j=1}^{149} \bigl|\bigl| \mathbf{Y}_{\bullet,j,\bullet} - f(\mathbf{X}_{\bullet,j,\bullet}; \boldsymbol{\alpha}_{j,\bullet}) \bigr|\bigr|_2^2,
\end{equation}
where, as in the last subsection, $N_{CF}$ denotes the number of CFs in $\mathbf{X}$ and $\mathbf{Y}$; we refer to this objective as ``$L_2$ Error''.
Our penalized estimation problem is defined in the following manner:
\begin{gather}
\begin{gathered}
\widehat{\boldsymbol{\alpha}} = \underset{\boldsymbol{\alpha}}{\argmin} \ L_2^{\boldsymbol{\alpha}}(\mathbf{Y}, F(\mathbf{X})) + \lambda_1 \bigl|\bigl| \boldsymbol{\alpha} \bigr|\bigr|_1 + \lambda_2 \sum_{l=1}^p \bigl|\bigl| \nabla^d (\boldsymbol{\alpha}_{\bullet,l}) \bigr|\bigr|_1, \\
\text{s.t. } 0 \leq \alpha_{j,l} \leq 1, \ 0 \leq \sum_{l=1}^p \alpha_{j,l} \leq 1, \ \forall j,l,
\end{gathered}
\label{eq:min_problem}
\end{gather}
where $\lambda_1, \lambda_2 \geq 0$ and $d \in \mathbb{N}$ signify tuning parameters.
The differencing order $d$ is used to specify a given level of smoothness in the spatial $\boldsymbol{\alpha}$ trend estimates because the $\sum_{l=1}^p \bigl|\bigl| \nabla^d (\boldsymbol{\alpha}_{\bullet,l}) \bigr|\bigr|_1$ term in the above minimization problem encourages $\boldsymbol{\alpha}$ trends that have $d$th order discrete derivatives close to 0 (i.e.\ that are piecewise polynomials of order $d-1$).
In addition, careful selection of $\lambda_1$ and $\lambda_2$ is required to obtain an adequate model fit.
Unfortunately, this is a constrained optimization problem with a multivariate output and there are not any obvious ways to minimize such an objective without resorting to general-purpose optimizers.
Therefore, in all our experiments, we use the L-BFGS-B algorithm \citep{byrd1995limited} to fit the above model.

\subsection*{Jaccard Similarity}
While the model described above has computational and statistical appeal, in engineering applications it is mostly interesting to know the high-frequency amino acid predictions; however, our penalized objective function focuses attention on the complete substitution profiles and not exclusively the high-frequency amino acids.
To provide a metric more closely aligned with antibody engineering goals, we utilize the Jaccard similarity metric, which can be used to measure differences between predicted and observed high-frequency amino acid sets.
Sets of high-frequency amino acids are defined at each position by a minimum frequency cutoff $t$; Jaccard similarities are then computed between the observed and predicted sets and averaged across each CF and AHo position in the dataset.

The Jaccard similarity metric \citep{jaccard1912distribution} measures the similarity between two finite sets.
Specifically, for any sets $A$ and $B$, the similarity metric $J(A,B)$ is defined as the ratio of the intersection size $|A \cap B|$ to the union size $|A \cup B|$.
It has these properties: $0 \leq J(A,B) \leq 1$; $J(A,B)=1$ when $A=B$ and $J(A,B)=0$ when $A \cap B = \emptyset$ (empty set).
To formally establish our use of Jaccard similarity, we define the following notation.
Let $\mathcal{Y}_{i,j} = \{y \in \mathbf{Y}_{i,j,\bullet} \mid y \geq t\}$ represent the set of amino acid frequencies at AHo position $j$ for CF $i$ that has observed frequencies greater than or equal to the cutoff $t$ and denote $\boldsymbol{\mathcal{Y}} \equiv \{ \mathcal{Y}_{i,j} \}$ for $i = 1, ..., N_{CF}$ and $j = 1, ..., 149$.
We define $\widehat{\mathcal{F}}^{\mathbf{X}}_{i,j}$ and $\widehat{\boldsymbol{\mathcal{F}}}^{\mathbf{X}} \equiv \{ \widehat{\mathcal{F}}^{\mathbf{X}}_{i,j} \}$ to be the analogous quantities for the predicted amino acid frequencies.
If we let $\mathcal{A}(\mathcal{Y}')$ denote a function that accepts as input an amino acid frequency set $\mathcal{Y}'$ (i.e.\ $\mathcal{Y}_{i,j}$ or $\widehat{\mathcal{F}}^{\mathbf{X}}_{i,j}$) and outputs the corresponding set of amino acid identities, then our Jaccard similarity objective can be written as:
\begin{equation}
J_t^{\boldsymbol{\alpha}} \equiv J_t^{\boldsymbol{\alpha}}(\mathbf{Y}, F(\mathbf{X})) = \frac{1}{149 \cdot N_{CF}} \sum_{i=1}^{N_{CF}} \sum_{j=1}^{149} J\bigl(\mathcal{A}(\mathcal{Y}_{i,j}), \mathcal{A}(\widehat{\mathcal{F}}^{\mathbf{X}}_{i,j})\bigr),
\end{equation}
which is referred to as the ``Jaccard Similarity'' objective.
We can define a penalized Jaccard estimation problem by substituting $-J_t^{\boldsymbol{\alpha}}(\mathbf{Y}, F(\mathbf{X}))$ for $L_2^{\boldsymbol{\alpha}}(\mathbf{Y}, F(\mathbf{X}))$ in Equation \eqref{eq:min_problem}.
Jaccard similarity optimization is difficult using derivative-based optimization because of its discrete nature, so we use a smooth approximation of the aforementioned metric for model fitting in our experiments (see Supplementary subsection Smoothed Jaccard Similarity).

\subsection*{Forward Stepwise Selection}
We devise a forward stepwise selection procedure to help us determine the combination of external profiles that best predict the outcome of interest, which can be penalized $L_2$ Error or Jaccard Similarity.
In this procedure, we initially try all possible external profiles in the model separately and determine the best fit using 5-fold cross-validation.
We cache the best model from the initial step and continue fitting models with two external profiles; the first external profile is fixed to be the best profile from the previous round and the second profile can be any possible remaining external profile.
We continue this iterative scheme until we reach a prespecified limit on the number of external profiles allowed in $\mathbf{X}^*$.
It is important to note that to ease computation, we perform forward selection using the unpenalized variants of our models.
Even though this procedure is greedy and not as thorough as all-subsets selection, we believe this technique provides the best trade-off between accuracy and efficiency.
We provide the implementation of our stepwise procedures at \url{https://github.com/krdav/SPURF}.

\subsection*{Inference Pipeline}
We apply a 80\%/20\% training/test split to the model fitting dataset described above.
We first run the forward stepwise selection procedure with a maximum profile limit of five to approximately determine the best profile groupings starting with a single profile and ending with a group of five profiles.
Using the profile groupings from the previous step, we fit the penalized version of the model and use 5-fold cross-validation to obtain estimates of the relevant tuning parameters, which consist of the lasso penalty weights $\lambda_1$, $\lambda_2$ and the differencing order $d$; note that we report unpenalized performance estimates when we run cross-validation.
After we determine the optimal tuning parameters via cross-validation, we fit the penalized model using the entire training portion of the model fitting dataset and the best tuning parameters and cache the resulting parameter estimates of $\boldsymbol{\alpha}$.
Once we obtain the estimates of $\boldsymbol{\alpha}$ from the penalized model, we can use them to compute the chosen performance metric on the testing portion of the model fitting dataset and any other validation dataset of interest.

\section*{Results}
As described in the methods (the Inference Pipeline subsection), we first need to infer the best profile groupings to use in penalized model fitting.
To determine these groupings, we run the forward stepwise selection procedure for both the $L_2$ error function and the smoothed Jaccard objective function with a frequency cutoff $t = 0.2$ (\autoref*{table:stepwise}).
For both objective functions, the forward selection path is the same until $\mathbf{X}^* = \bigl\{\widehat{\mathbf{X}}_{\naiveAA}, \widehat{\mathbf{X}}_{\vgene}, \widehat{\mathbf{X}}_{\neut}, \widehat{\mathbf{X}}_{\vsubgrp}\bigr\}$.
For the $L_2$ loss function, model performance is the best when $\mathbf{X}^* = \bigl\{\widehat{\mathbf{X}}_{\naiveAA}, \widehat{\mathbf{X}}_{\vgene}, \widehat{\mathbf{X}}_{\neut}, \widehat{\mathbf{X}}_{\vsubgrp}\bigr\}$ even though there are diminishing returns for using profiles beyond $\mathbf{X}^* = \bigl\{\widehat{\mathbf{X}}_{\naiveAA}, \widehat{\mathbf{X}}_{\vgene}\bigr\}$.
In a similar fashion, the Jaccard similarity estimates tend to be highest when $\mathbf{X}^* = \bigl\{\widehat{\mathbf{X}}_{\naiveAA}, \widehat{\mathbf{X}}_{\vgene}\bigr\}$, despite the almost identical model performance from just using $\mathbf{X}^* = \bigl\{\widehat{\mathbf{X}}_{\naiveAA}\bigr\}$.
For the subsequent penalized model fitting step, we choose to evaluate the $\bigl\{\widehat{\mathbf{X}}_{\naiveAA}, \widehat{\mathbf{X}}_{\vgene}, \widehat{\mathbf{X}}_{\neut}\bigr\}$ and $\bigl\{\widehat{\mathbf{X}}_{\naiveAA}, \widehat{\mathbf{X}}_{\vgene}, \widehat{\mathbf{X}}_{\neut}, \widehat{\mathbf{X}}_{\vsubgrp}\bigr\}$ profile groupings with the $L_2$ objective and $\bigl\{\widehat{\mathbf{X}}_{\naiveAA}\bigr\}$ and $\bigl\{\widehat{\mathbf{X}}_{\naiveAA}, \widehat{\mathbf{X}}_{\vgene}\bigr\}$ with the smoothed Jaccard similarity objective.
\begin{table}[ht!]
\centering
\begin{tabular}{lcccccc}
 & & & & & \\[-9pt]
Objective Function & \multicolumn{6}{c}{Unregularized CV} \\
\hline
 & & & & & & \\[-10pt]
\multirow{2}{*}{$L_2$ Error} & $\varnothing$ & $\widehat{\mathbf{X}}_{\naiveAA}$ & $\widehat{\mathbf{X}}_{\vgene}$ & $\widehat{\mathbf{X}}_{\neut}$ & $\widehat{\mathbf{X}}_{\vsubgrp}$ & $\widehat{\mathbf{X}}_{\naiveAAclust\text{-}5}$ \\
 & 0.110 & 0.0542 & 0.0459 & 0.0456 & 0.0455 & 0.0456 \\
\hline
 & & & & & & \\[-10pt]
Jaccard Similarity & $\varnothing$ & $\widehat{\mathbf{X}}_{\naiveAA}$ & $\widehat{\mathbf{X}}_{\vgene}$ & $\widehat{\mathbf{X}}_{\neut}$ & $\widehat{\mathbf{X}}_{\vsubgrp}$ & $\widehat{\mathbf{X}}_{\naiveAAclust\text{-}85}$ \\
($t = 0.2$) & 0.9170 & 0.9322 & 0.9324 & 0.9323 & 0.9319 & 0.9318
\end{tabular}
\caption{Results of forward stepwise selection on our $L_2$ and smooth Jaccard objective functions.
The performance estimates shown in the table are obtained using 5-fold cross-validation.
Going from left to right, each column represents the best profile addition into $\mathbf{X}^*$ with the associated CV performance estimate.
For Jaccard, we fit using the smooth Jaccard objective, but report exact Jaccard similarity estimates, both using frequency cutoff $t = 0.2$.
Note that we fix the prespecified limit on the number of external profiles allowed in $\mathbf{X}^*$ to be 5.
$\varnothing$ represents the model using only the input sequence.
}
\label{table:stepwise}
\end{table}

We now use the approximate profile groupings obtained from the forward stepwise selection procedure to fit our regularized models.
The penalized estimation problem has additional tuning parameters that must be determined.
In our experiments, we cross-validate over penalty parameters; $\lambda_1, \lambda_2 = 10^{-7}, 5.05 \times 10^{-6}, 10^{-5}$; the differencing order, $d = 1, 2, 3$; and the two profile groupings specified above for both the $L_2$ error and Jaccard similarity objectives.
The best regularized $L_2$ model uses $\mathbf{X}^* = \bigl\{\widehat{\mathbf{X}}_{\naiveAA}, \widehat{\mathbf{X}}_{\vgene}, \widehat{\mathbf{X}}_{\neut}, \widehat{\mathbf{X}}_{\vsubgrp}\bigr\}$, while the best regularized Jaccard model utilizes $\mathbf{X}^* = \bigl\{\widehat{\mathbf{X}}_{\naiveAA}\bigr\}$ (\autoref*{table:cv}).
In summary, using many external profiles is important for predicting the complete substitution profiles, while the inferred naive sequence is the only external profile deemed useful for our model to accurately predict the observed high-frequency amino acids (where high-frequency is defined by being at least 20\% of the observed amino acids).

\begin{figure}[ht!]
\centering
\includegraphics[width=\textwidth, trim=0 140 0 0, clip]{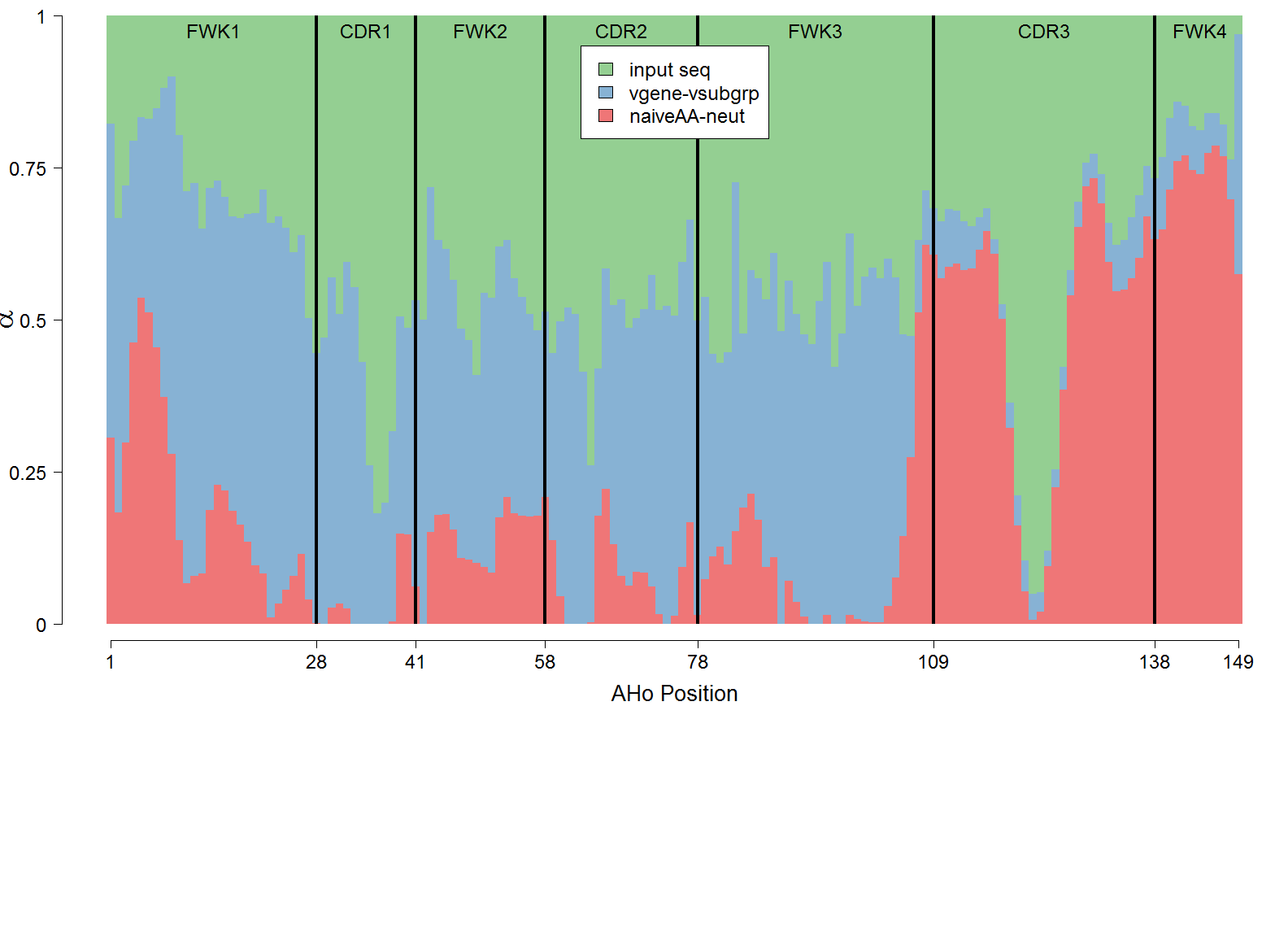}
\caption{A stacked barplot of the estimated parameter values of $\boldsymbol{\alpha}$ from the best regularized $L_2$ model.
For convenience, we aggregate the estimates of $\boldsymbol{\alpha}$ associated with $\widehat{\mathbf{X}}_{\vgene}$ and $\widehat{\mathbf{X}}_{\vsubgrp}$ (blue) and with $\widehat{\mathbf{X}}_{\naiveAA}$ and $\widehat{\mathbf{X}}_{\neut}$ (red).
The black vertical lines represent the boundaries between the different CDRs and FWKs.
}
\label{fig:alpha_profile_plot_collapsed}
\end{figure}

\begin{figure}[ht!]
\centering
\includegraphics[width=\textwidth]{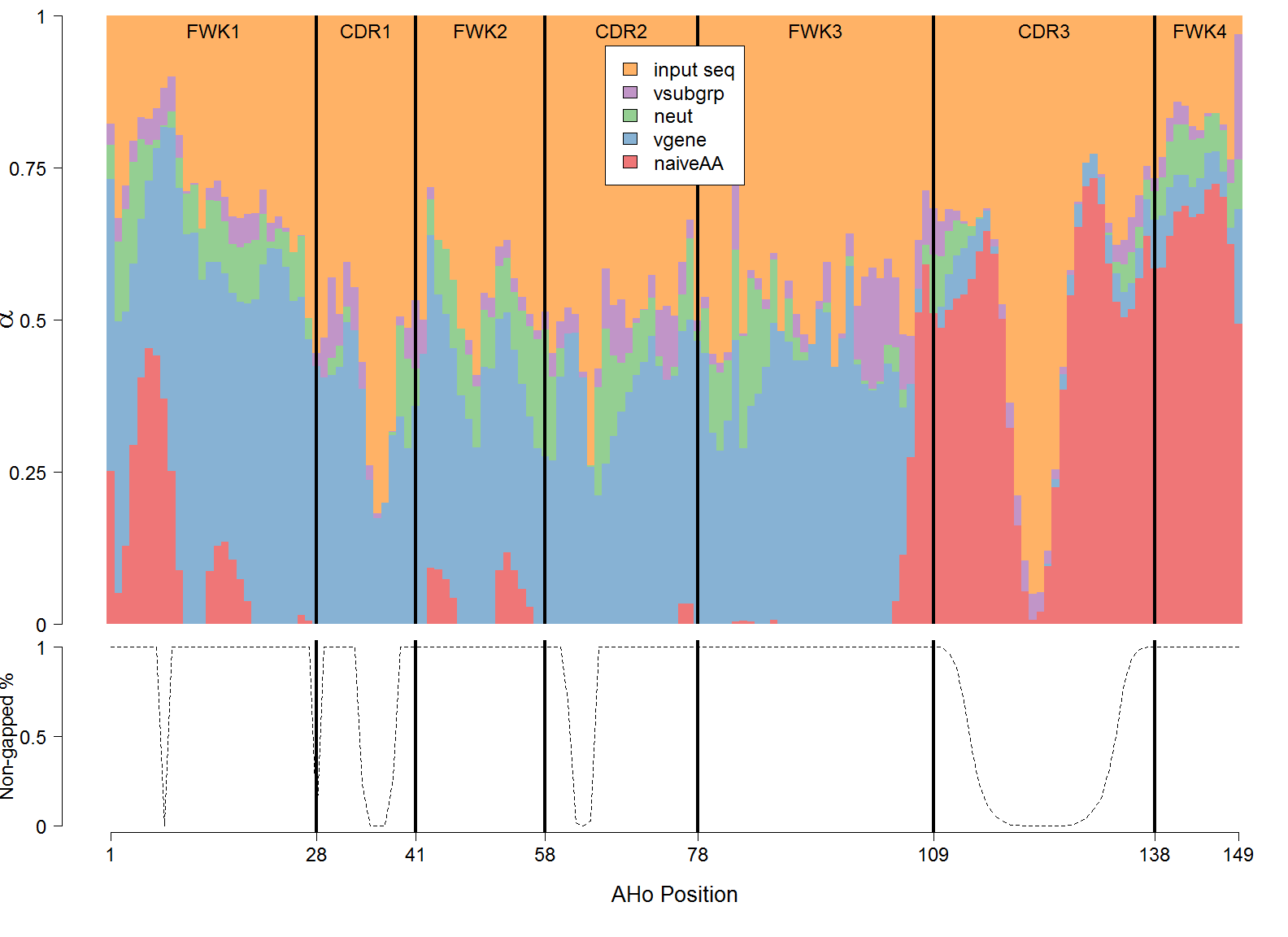}
\caption{A stacked barplot of the estimated parameter values of $\boldsymbol{\alpha}$ from the best regularized $L_2$ model.
The black vertical lines represent the boundaries between the different CDRs and FWKs.
Due to the AHo antibody numbering used \citep{honegger2001yet}, some positions are assigned to a gap character (an AHo position that does not map to a sequence position).
The percentage of CFs that are not assigned to gap characters is shown in the bottom plot for each AHo position.
The input sequence is heavily weighted in regions with high gap percentages because of the standard lasso penalty included in our model.
The conserved Tryptophan amino acid is observed as a spike in the $\widehat{\mathbf{X}}_{\vgene}$ and $\widehat{\mathbf{X}}_{\naiveAA}$ profile weights following the end of CDR1 (position 43 in the AHo scheme).
The conserved Cysteine amino acid that defines the beginning of CDR3 is not readily observed, presumably because this is invariant in all profiles.
Generally, the input sequence has less weight in CDR3 and FWK4, which indicates that there is some conservation during affinity maturation.
Beyond CDR3 and FWK4, there is a general trend that the input sequence has higher weight in the CDRs than in the FWKs, which suggests that there is a higher level of conservation in the FWKs than in the CDRs during affinity maturation.
A more surprising observation is the spike in the $\widehat{\mathbf{X}}_{\vgene}$, $\widehat{\mathbf{X}}_{\vsubgrp}$, and $\widehat{\mathbf{X}}_{\neut}$ weights at AHo position 83 near the beginning of FWK3 (the ``outer'' loop); this could indicate a conserved position not previously described.
}
\label{fig:alpha_plot}
\end{figure}

In addition to predictive performance, we are also interested in understanding how the estimated parameter weights from our best regularized $L_2$ model vary across the different external profiles in $\mathbf{X}^*$ and antibody regions.
For convenience, we aggregate the estimates of $\boldsymbol{\alpha}$ associated with the V gene ($\widehat{\mathbf{X}}_{\vgene}$ and $\widehat{\mathbf{X}}_{\vsubgrp}$) and with the full naive sequence ($\widehat{\mathbf{X}}_{\naiveAA}$ and $\widehat{\mathbf{X}}_{\neut}$) as these sets of profiles are intuitively similar (\autoref*{fig:alpha_profile_plot_collapsed}); the V-gene and V-subgroup profiles are both derived by averaging over different IMGT V germline gene labeling schemes and the simulated S5F neutral substitution profiles originate from the CF-specific inferred naive sequence.
Antibody heavy chain (and light chain) sequences can be partitioned into framework regions (FWKs) and complementarity-determining regions (CDRs) by the AHo definitions \citep{honegger2001yet}; the BCR binding affinity is largely determined by the CDRs (especially by the heavy chain CDR3), while the FWKs encode the structural constraints of the BCR and thus can be strongly conserved \citep{tomlinson1995structural}.
The $\widehat{\mathbf{X}}_{\vgene}$ and $\widehat{\mathbf{X}}_{\vsubgrp}$ profiles are extremely important for prediction at FWK1-FWK3, which is not surprising as V germline genes extend from the FWK1 to the beginning of the CDR3.
In contrast, the $\widehat{\mathbf{X}}_{\naiveAA}$ and $\widehat{\mathbf{X}}_{\neut}$ external profiles are heavily weighted in the CDR3 and FWK4; this result is also intuitive because the CDR3 is highly variable across CFs as it is a strong determinant of antigen-binding specificity, the $\widehat{\mathbf{X}}_{\naiveAA}$ and $\widehat{\mathbf{X}}_{\neut}$ profiles are our only CF-specific sources of external information, and the V gene specific profiles cannot provide any information beyond the end of the V gene.
Furthermore, the FWKs have, on average, more support from the external profiles compared to the CDRs, which is consistent with our understanding of antibody biochemistry as the FWKs are structurally constrained and thus need to be more conserved compared to the more flexible CDRs.
We note that the middle of the CDR3 has artificially low estimates of $\boldsymbol{\alpha}$ because most of the AHo positions in the CDR3 have only a few or no defined sequence positions in the dataset (\autoref*{fig:alpha_plot}).

While our penalized modeling framework allows for easy interpretation of the parameter estimates, ultimately the quality of the $\boldsymbol{\alpha}$ estimates is determined by their performance on independent test datasets.
Specifically, we compute the $L_2$ error ($L_2^{\boldsymbol{\alpha}}$) and Jaccard similarity ($J_{0.2}^{\boldsymbol{\alpha}}$) between the predicted and observed profiles associated with both the testing portion of the model fitting dataset and the Briggs validation dataset (\autoref*{table:pred}); we remind readers that these predictions are made based on the subsampled (i.e.\ single-sequence) profiles in the aforementioned datasets and compared to the corresponding actual substitution profiles through the $L_2^{\boldsymbol{\alpha}}$ and $J_{0.2}^{\boldsymbol{\alpha}}$ performance metrics (\autoref*{fig:model_overview}).
Our model improves upon the ``baseline'' prediction performance, where ``baseline'' refers to predictions made using only the input sequence (i.e.\ model predictions with all parameter values of $\boldsymbol{\alpha}$ set to 0).
\begin{table}[ht!]
\centering
\begin{tabular}{llcc}
 & & \\[-9pt]
Objective Function & Model Type & Model fitting: test & Briggs \\
\hline
 & & & \\[-9pt]
\multirow{2}{*}{$L_2$ Error} & Best & 0.0492 & 0.0511 \\
 & Baseline & 0.114 & 0.129 \\
\hline
 & & & \\[-9pt]
Jaccard Similarity & Best & 0.9289 & 0.9227 \\
($t = 0.2$) & Baseline & 0.9156 & 0.9053
\end{tabular}
\caption{The model performance results from predicting on independent datasets.
We provide results for both the testing portion of the model fitting dataset and the Briggs validation dataset.
Note that the term ``baseline'' refers to predictions made using only the input sequence (i.e.\ model predictions with all parameter values of $\boldsymbol{\alpha}$ set to 0).
}
\label{table:pred}
\end{table}

\begin{figure}[ht!]
\centering
\includegraphics[width=\textwidth, trim=0 15 0 0, clip]{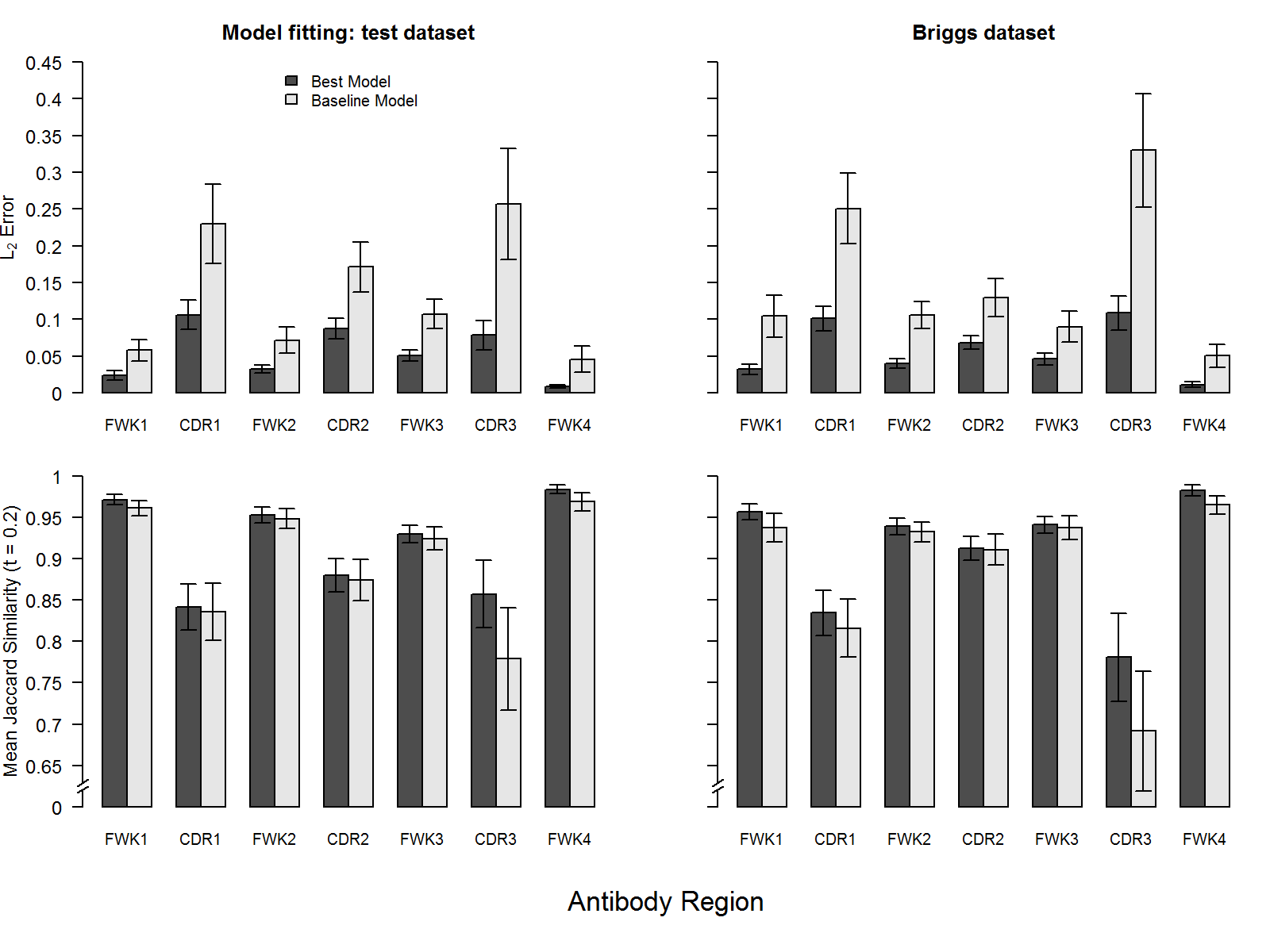}
\caption{The model performance results across the different antibody regions on the model fitting test dataset and the Briggs validation dataset.
In these plots, we compare the performances from our best models to the baseline predictive performances using only the input sequence (i.e.\ model predictions with all parameter values of $\boldsymbol{\alpha}$ set to 0).
The error bars show bootstrap standard errors.
}
\label{fig:pred_fwkcdr}
\end{figure}

In addition, we also want to know how well our model performs in the different antibody regions (i.e.\ FWKs/CDRs).
To answer this question, we compute the same metrics as shown in \autoref*{table:pred} for the different FWKs and CDRs (\autoref*{fig:pred_fwkcdr}).
To provide some insight into the variability of the model performance estimates in the different regions, we calculate bootstrap standard errors, which are expressed as error bars in \autoref*{fig:pred_fwkcdr}.

We see that our substitution profile prediction model performs well in the CDRs relative to the baseline model.
This is an important finding because antigen binding is largely determined by the sequence segments in the CDRs, and especially CDR3.
In fact, our models seem to provide the greatest improvement in performance in the CDR3, which is also the hardest region to predict because it has the highest amount of sequence variability.
Another important takeaway is that the prediction performance is better in FWKs than CDRs, which is presumably because FWKs have lower variance and are more conserved compared to CDRs.
In summary, our prediction models are able to systematically integrate different data sources to make better predictions of the per-site amino acid compositions in CFs.

Our model also improves the prediction of the highest-frequency amino acid at a given position, referred to here as the mode (\autoref*{table:mode_pred}).
Indeed, the counts in the bottom-left cells (cases where the model is correctly predicting the actual mode given an incorrect input sequence amino acid) are larger than the counts in the top-right cells (vice-versa).
In addition, the input sequence amino acids that are not the true modes but correctly predicted by the model to be the actual modes are all germline reversions, which is consistent with the $\widehat{\mathbf{X}}_{\naiveAA}$ profile being heavily weighted in our prediction model (\autoref*{fig:alpha_profile_plot_collapsed}).
In the opposite case, where the input sequence amino acid is correct but the model prediction is wrong, all the counts consist of germline predictions as well.
In summary, many of the mode predictions are just germline reversions and, in fact, most of these predictions are to the true modes (i.e.\ the actual highest-frequency amino acids); however, most of the input sequence amino acids are the true modes already ($\approx$ 99\%).
\begin{table}[ht!]
\minipage{0.5\textwidth}
\begin{subtable}{\textwidth}
\centering
\begin{tabular}{cc|C{1.45cm}C{1.45cm}}
\hline
 & & & \\[-9pt]
 & & \multicolumn{2}{c}{Correct SPURF} \\
 \multicolumn{2}{c|}{germline\,$\mid$\,non-germline} & \multicolumn{2}{c}{Mode Prediction?} \\
 & & & \\[-9pt]
 & & Yes & No \\
\hline
 & & & \\[-9pt]
Is input amino & Yes & 10,473\,$\mid$\,465 & 156\,$\mid$\,0 \ \ \ \ \\
acid the mode? & No &\hspace{3.75mm}349\,$\mid$\,0 & 170\,$\mid$\,395
\end{tabular}
\caption{Model fitting: test dataset}
\label{table:test_mode_pred}
\end{subtable}
\endminipage
\minipage{0.5\textwidth}
\begin{subtable}{\textwidth}
\centering
\begin{tabular}{cc|C{1.45cm}C{1.45cm}}
\hline
 & & & \\[-9pt]
 & & \multicolumn{2}{c}{Correct SPURF} \\
 \multicolumn{2}{c|}{germline\,$\mid$\,non-germline} & \multicolumn{2}{c}{Mode Prediction?} \\
 & & & \\[-9pt]
 & & Yes & No \\
\hline
 & & & \\[-9pt]
Is input amino & Yes & 10,541\,$\mid$\,376 & 178\,$\mid$\,1\ \ \ \ \ \\
acid the mode? & No &\hspace{3.75mm}474\,$\mid$\,0 & 196\,$\mid$\,393
\end{tabular}
\caption{Briggs dataset}
\label{table:juno_mode_pred}
\end{subtable}
\endminipage
\caption{Mode prediction results from both the testing portion of the model fitting dataset and the Briggs dataset.
For each CF and AHo position in a given dataset, we determine whether the predicted mode (i.e.\ highest-frequency amino acid) from our best model is the same as the actual mode.
Results are aggregated based on whether or not the input sequence has the correct mode.
At the left side of the vertical bar ($\mid$) is the count for the germline predicted modes (i.e.\ situations when the predicted amino acid mode is the naive sequence amino acid) and at the right side is the count for the non-germline predicted modes (vice-versa).
}
\label{table:mode_pred}
\end{table}

\begin{figure}[ht!]
\centering
\includegraphics[width=\textwidth]{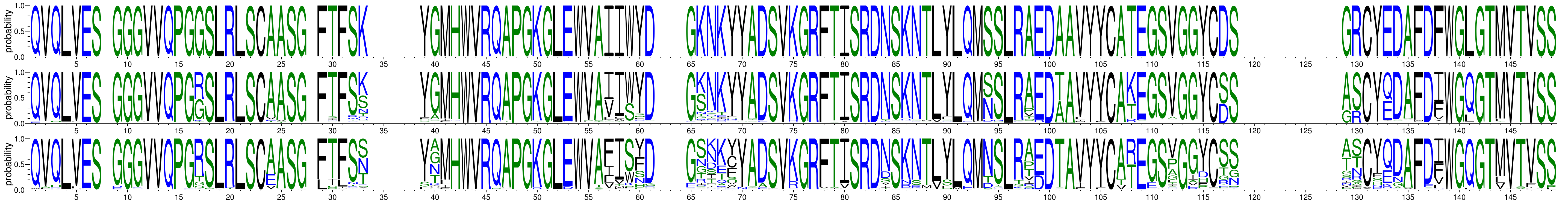}
\caption{A logo plot displaying the input sequence, predicted profile, and true profile (ordered from top to bottom) for an arbitrary CF in the Briggs dataset.
The logos are plotted using AHo numbers (1-149) and AHo positions undefined in the sequence are shown as empty columns.
The predicted profile (middle) captures much of the amino acid composition information associated with the full profile (bottom).
}
\label{fig:juno_logo}
\end{figure}

The in-sample and out-of-sample prediction performances demonstrate that our SPURF inference pipeline is able to obtain accurate and robust estimates of $\boldsymbol{\alpha}$.
Specifically, prediction performance is consistently similar but slightly worse when comparing the Briggs dataset to the model fitting test set, which likely reflects two things: 1) the median number of sequences per CF in the Briggs set is lower than in the test set (\autoref*{table:ds_sumstats}) and 2) the model fitting dataset is sampled from the same donors as the dataset for cross-validation.
Regardless, the differences between the test and Briggs datasets are small, which provides evidence in support of our model performance estimates.
Subjective assessments of the inferred substitution profiles coincide with our description of the $L_2$ error metric, namely that fine-grained amino acid substitution information is captured by SPURF (\autoref*{fig:juno_logo}).

\begin{figure}[ht!]
\centering
\includegraphics[width=\textwidth]{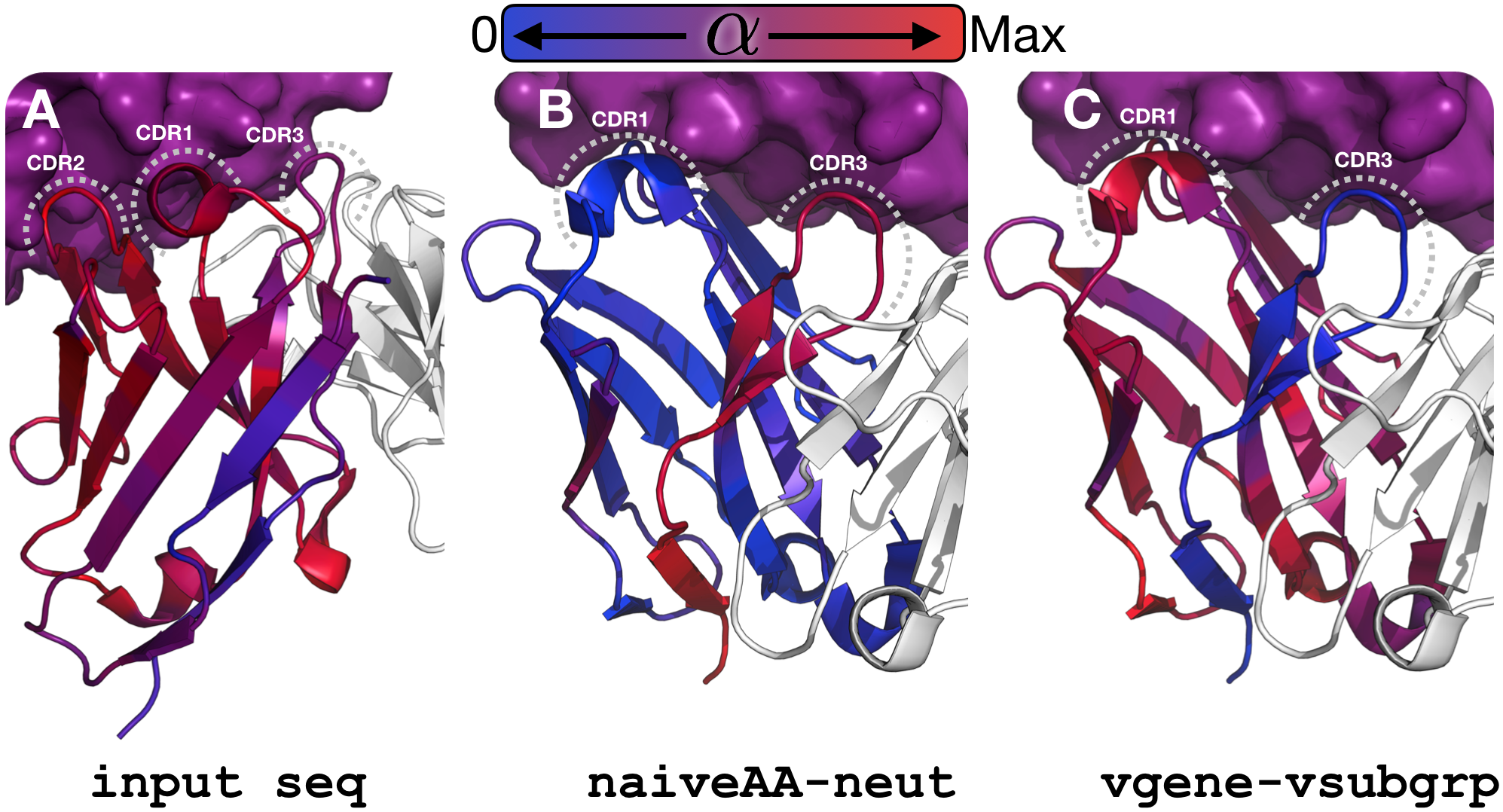}
\caption{Positional profile weights $\boldsymbol{\alpha}$ mapped to an antibody protein structure (PDB: 5X8L).
The antigen (PD-L1) appears as a purple surface at the top of the images, the light chain appears in white cartoon, and the heavy chain is displayed using a blue to red color gradient; the grey dashed lines mark the CDR loops.
The color gradient represents the possible values of profile weights in $\boldsymbol{\alpha}$ and goes from blue at a zero weight to red at the maximum weight for the profile.
The display in panels \textbf{B} and \textbf{C} is rotated relative to panel \textbf{A} to better show results for CDR1 and CDR3; as a consequence, the CDR2 loop is hidden behind the CDR1.
Panel \textbf{A} shows that the input sequence has high weight at the CDR1 and CDR2, panel \textbf{B} illustrates that the naive sequence and the neutral substitution profile have high weight at the CDR3 and FWK4, and panel \textbf{C} demonstrates that the V gene and V subgroup profiles are highly weighted in parts of the CDR1 but more generally in the FWKs, especially at the heavy and light chain interface.
}
\label{fig:alpha_on_protein}
\end{figure}

\begin{figure}[ht!]
\centering
\includegraphics[width=0.9\textwidth]{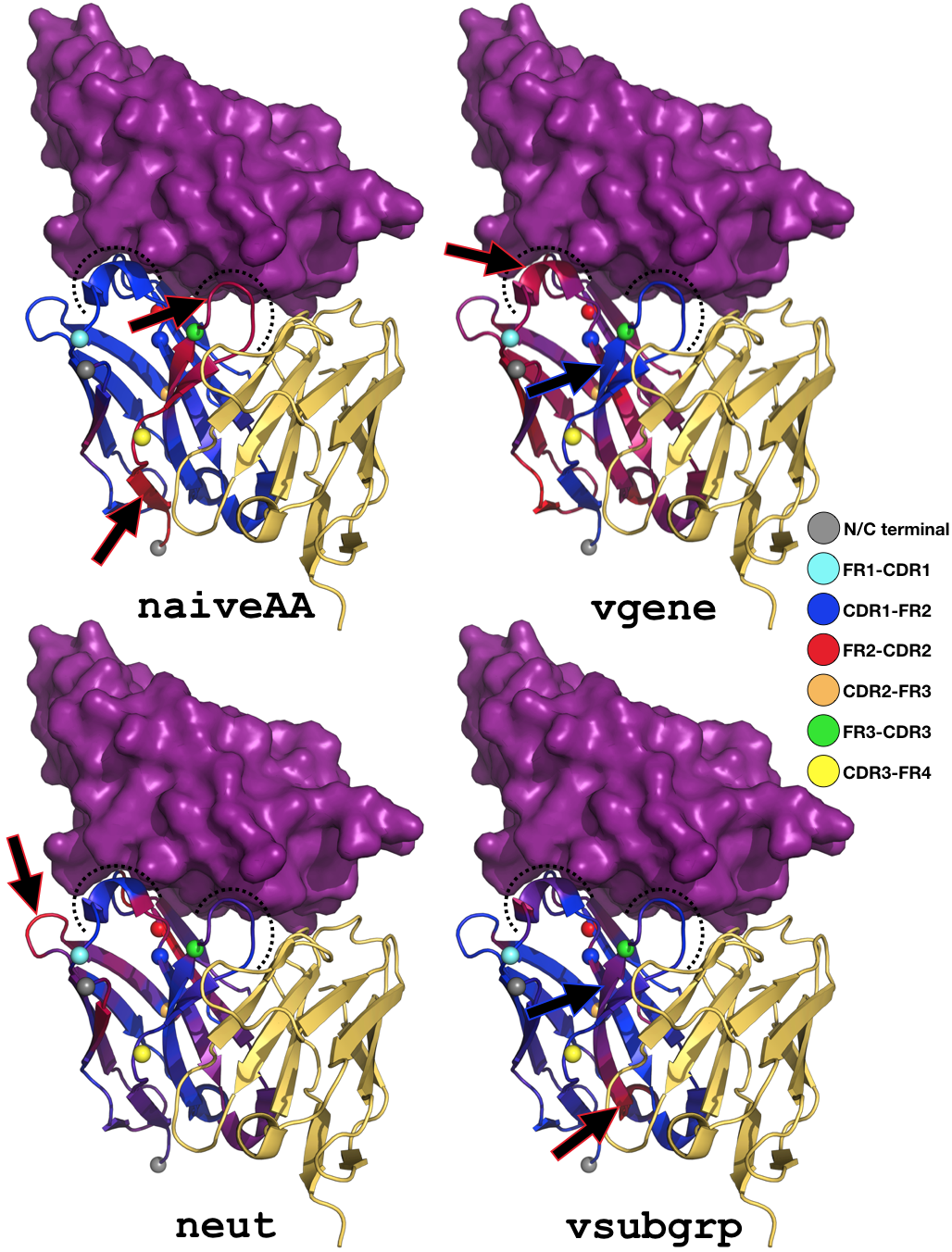}
\caption{Positional profile weights $\boldsymbol{\alpha}$ mapped to an antibody protein structure (PDB: 5X8L).
The antigen (PD-L1) appears as a purple surface at the top of the images, the light chain appears in yellow cartoon, and the heavy chain is displayed using a blue to red color gradient.
The color gradient represents the possible values of profile weights in $\boldsymbol{\alpha}$ and goes from blue at a zero weight to red at the maximum weight for the profile.
The black dashed lines mark the CDR loops; note that the CDR2 loop is hidden behind the CDR1.
The colored balls represent the AHo-defined FWK/CDR boundaries.
The black arrows indicate regions of high profile weight.
The $\widehat{\mathbf{X}}_{\naiveAA}$ profile is heavily weighted in CDR3 and FWK4.
The $\widehat{\mathbf{X}}_{\vgene}$ profile weighting is fairly even from FWK1 through FWK3; it spikes slightly in CDR1 and completely disappears beyond FWK3, which is expected as the V-D junction region starts past the end of FWK3.
The $\widehat{\mathbf{X}}_{\neut}$ profile weighting is fairly even across sites but spikes near the beginning of FWK3 (the ``outer'' loop).
The $\widehat{\mathbf{X}}_{\vsubgrp}$ profile weighting is distributed similarly to that of the $\widehat{\mathbf{X}}_{\vgene}$ profile with the exception of a spike at the end of FWK3 (i.e.\ at the heavy and light chain interface).
}
\label{fig:alpha_on_protein_all4}
\end{figure}

The SPURF model setup produces interpretable and meaningful profile weights (\autoref*{fig:alpha_on_protein}; per-profile decomposition in \autoref*{fig:alpha_on_protein_all4}).
The input sequence is strongly weighted in the CDRs, indicating that substitutions in these regions are both specific and conserved within the CF and, therefore, cannot easily utilize the information from other CFs.
The weight on the V gene specific profiles spikes at CDR1 and at the end of FWK3, which is at the heavy and light chain interface.
We note that, as expected, the weight on the V gene specific profiles is minimal downstream of FWK3 as this is the end of the V gene and the beginning of the V-D junction region.
As such, nothing prevents the V gene profiles from having a high weight downstream of FWK3, but the model framework has chosen these meaningful weights without any manual interference.
We ascribe this shrinkage feature of the weights to the standard lasso penalty built into SPURF.
The profiles that are derived from the inferred naive sequence ($\widehat{\mathbf{X}}_{\naiveAA}$, $\widehat{\mathbf{X}}_{\neut}$) take up the missing weight of the V gene profiles as these are highly weighted in the CDR3 and FWK4.

\section*{Discussion}
In this paper, we present SPURF, a statistical framework for predicting CF-specific amino acid frequency profiles from single input BCR sequences by leveraging multiple sources of external information.
We use standard and spatial lasso penalties to prevent our model from overfitting and obtain sparse, interpretable estimates of the profile weights, expressed by an $\boldsymbol{\alpha}$ matrix.
The spatial lasso penalizes extreme differences between spatially-adjacent profile weights, while the standard lasso penalties promote simpler models by shrinking parameter values in $\boldsymbol{\alpha}$ to 0 if the associated external profiles are not useful predictors.
We show that our method not only performs well on the held-out (test) portion of our model fitting dataset but also provides accurate predictions on the Briggs external validation dataset.
Indeed, we did not obtain the Briggs validation dataset until after we ran our model inference pipeline on the model fitting dataset.

Our work can be seen as a prediction-based extension of the work of \citet{sheng2017gene} and \citet{Kirik2017-bc}.
This previous work illustrates that amino acid substitution profiles differ between germline genes, a finding supported by the context specificity of somatic hypermutation \citep{cui2016model}.
In our work, we provide a prediction algorithm that takes a single BCR sequence from a clonal family as input and outputs a CF-specific substitution profile estimate for the whole VDJ region.

We believe that this work will be a useful tool for antibody engineering in situations when it is important to maintain antibody binding affinity to the same epitope.
The predicted profiles from SPURF can be used to choose the sites that are most tolerable for mutagenesis and the substitutions that are most likely to maintain binding specificity; as such, this information can be used to engineer antibodies with better biophysical properties.

To our knowledge, SPURF is the first prediction algorithm for B cell CF substitution profiles.
There are many possible extensions.
In our SPURF inference pipeline, we subsample single BCR sequences from CFs to use as model input; unfortunately, this means that our modeling analysis is conditional on a dataset that does not account for the variability associated with the subsampling process.
One obvious means of fixing the above problem is to draw multiple subsamples from each CF and treat these multiple ``observations'' per CF within a dataset as a clustered data or weighted least squares problem.
In addition, our model fitting dataset consists of only the largest CFs because we need accurate CF-specific substitution profile estimates to serve as the ground truth.
This non-random sampling technique could potentially bias our analysis results; however, this appears unlikely given our model's performance on the external Briggs validation dataset.
Furthermore, our approach models per-site amino acid composition in a CF and accounts for interactions between sites only through the fusion lasso penalties.
It is well known from other protein studies that spatially-adjacent amino acid residues evolve jointly \citep{jones2011psicov,ekeberg2013improved}, presumably to maintain structural stability, or in the case of antibodies to stabilize the interface between heavy and light chains \citep{wang2009interactions}.
In the context of antibodies, residues in the FWKs have the potential to co-evolve (e.g.\ FWK residues flanking the CDRs could co-evolve to stabilize the stem leading to the more flexible CDRs).
Thus, figuring out how to incorporate more detailed interaction effects in our model is an important avenue for future research.

\section*{Acknowledgments}
We would like to thank Jason A.\ Vander Heiden and Steven H.\ Kleinstein for sharing post-processed data (dataset 1-4), Mikhail Shugay for sharing post-processed data (dataset 5), and Uri Hershberg for providing the ImmuneDB data (dataset 6).
We would also like to thank Juno Therapeutics, Inc.\ for providing and preparing the single cell dataset used as our external validation.
This research was supported by National Institutes of Health grants R01 GM113246, R01 AI12096, and U19 AI117891.
Amrit Dhar was supported by an NSF IGERT DGE-1258485 fellowship.
The research of Frederick Matsen was supported in part by a Faculty Scholar grant from the Howard Hughes Medical Institute and the Simons Foundation.

\bibliographystyle{apa}
\bibliography{main.bib}

\beginsupplement
\clearpage

\pagenumbering{arabic}
\renewcommand*{\thepage}{S-\arabic{page}}

\section*{Supplementary Materials}
\subsection*{Model Interpretation}
In this subsection, we provide statistical motivation for our penalized regression model, which can be interpreted as specifying an ensemble of multinomial logistic regression models at each AHo position.
We use some of the notation mentioned in the methods section and introduce new notation as needed.
We begin by describing the structure of the multinomial logistic regression models and then discuss how we perform model averaging with these component models to form the estimator $F(\mathbf{X})$ as stated in the methods section.
We conclude this subsection by showing that our regularized minimization problem can be characterized as a maximum a posteriori (MAP) inference problem.

Suppose we observe $M$ amino acids at the $j$th AHo position for the $i$th CF; for simplicity, we let $y_1, ..., y_M$ denote the observed amino acids.
We assume that $y_1, ..., y_M$ are drawn independently from a common multinomial distribution with 20 possible categories and define a logistic regression model for the substitution probabilities that does not include covariates.
The standard way to formulate such a model is as follows:
\begin{equation*}
\text{log}\biggl(\frac{\text{P}(y_m = c)}{\text{P}(y_m = 20)}\biggr) = \beta^{(c)}, \qquad \forall c \in \{ 1, ..., 20 \},
\end{equation*}
where $c$ indexes a particular amino acid, $\beta^{(20)} \equiv 0$, and $m = 1, ..., M$.
We can equivalently represent the model as:
\begin{equation*}
\text{P}(y_m = c) = \frac{\text{exp}(\beta^{(c)})}{\sum_{c'=1}^{20} \text{exp}(\beta^{(c')})}, \qquad \forall c \in \{ 1, ..., 20 \},
\end{equation*}
where $m = 1, ..., M$.
If we let $\widehat{p}_c$ denote the observed proportion of amino acid $c$ in the sample, then it is easy to show that maximizing the multinomial likelihood of the $M$ observations with respect to the $\beta^{(c)}$ parameters leads to the following parameter estimates:
\begin{equation*}
\widehat{\beta}^{(c)} = \text{log}\biggl(\frac{\widehat{p}_c}{\widehat{p}_{20}}\biggr), \qquad \forall c \in \{ 1, ..., 20 \},
\end{equation*}
which implies that:
\begin{equation*}
\widehat{\text{P}}(y_m = c) = \widehat{p}_c, \qquad \forall c \in \{ 1, ..., 20 \},
\end{equation*}
where $\widehat{\text{P}}(y_m = c)$ represents the logistic regression estimate of $\text{P}(y_m = c)$.
Therefore, this multinomial logistic regression model provides simple, intuitive estimates of the substitution probabilities.
While these logistic regression estimates may seem trivial, the underlying framework allows for easy integration of CF-specific and site-specific information into our model.

The above model considers the substitution probabilities at only one AHo position so one could fit this multinomial logistic regression model at each of the 149 AHo positions in a CF to obtain a complete substitution profile estimate.
In this paper, each CF-specific substitution profile estimate is a maximum likelihood estimate (MLE) obtained by fitting 149 multinomial regression models to the observed amino acid data in the CF.
Given that our proposed modeling procedure computes a per-site weighted average between the subsampled profiles $\mathbf{X}$ and all the external profile estimates $\mathbf{X}^*$, one can interpret this model $F(\mathbf{X})$ as defining an ensemble of multinomial logistic regression estimates at each AHo position.

To specify this relationship more clearly, we denote the likelihood of $\mathbf{Y}_{i,j,\bullet}$ as follows:
\begin{gather*}
\mathbf{Y}_{i,j,\bullet} \sim \text{MVN}\bigl(\boldsymbol{\mu}_{i,j,\bullet}, \sigma^2 \mathbb{I}_{20}\bigr),\\
\boldsymbol{\mu}_{i,j,\bullet} \equiv \sum_{l=1}^p \alpha_{j,l} \cdot \mathbf{X}^*_{i,j,\bullet,l} + \Bigl( 1 - \sum_{l=1}^p \alpha_{j,l} \Bigr) \cdot \mathbf{X}_{i,j,\bullet},
\end{gather*}
where $\text{MVN}$ means multivariate normal, $\boldsymbol{\mu}_{i,j,\bullet}$ signifies the mean vector of $\mathbf{Y}_{i,j,\bullet}$ and defines our ensemble model, $\sigma^2$ represents an unknown variance parameter, $\mathbb{I}_{20}$ symbolizes the $20 \times 20$ identity matrix, and $p$ represents the number of external profiles in $\mathbf{X}^*$.
Note that $\boldsymbol{\mu}_{i,j,\bullet}$ depends on the multinomial logistic regression estimates described previously as both $\mathbf{X}$ and $\mathbf{X}^*$ contain the MLE-based substitution profile estimates.
In addition, the form of $\boldsymbol{\mu}_{i,j,\bullet}$ relates to our previous definition of $F(\mathbf{X})$ by observing that $\boldsymbol{\mu}_{\bullet,j,\bullet} = f(\mathbf{X}_{\bullet,j,\bullet}; \boldsymbol{\alpha}_{j,\bullet})$.
We can also integrate the inequality constraints of the ensemble weights $\alpha_{j,l}$ into the likelihood function by including the indicator term $\mathbbm{1}_{\boldsymbol{\alpha}} \equiv \mathbbm{1}\{ 0 \leq \alpha_{j,l} \leq 1; \ 0 \leq \sum_{l=1}^p \alpha_{j,l} \leq 1; \ \forall j,l \}$.
The lasso penalties can be incorporated into our model through the use of sparsity-inducing prior distributions.

Specifically, the priors placed on $\boldsymbol{\alpha}$ can be represented as:
\begin{equation*}
\text{P}(\boldsymbol{\alpha}_{\bullet,l}) \propto \text{exp}\biggl(-\lambda_1 \bigl|\bigl| \boldsymbol{\alpha}_{\bullet,l} \bigr|\bigr|_1 - \lambda_2 \bigl|\bigl| \nabla^d (\boldsymbol{\alpha}_{\bullet,l}) \bigr|\bigr|_1\biggr), \qquad \forall l \in \{ 1, ..., p \},
\end{equation*}
where $\lambda_1, \lambda_2 \geq 0$ and $d \in \mathbb{N}$ are the same tuning parameters specified in the methods section \citep{park2008bayesian,kyung2010penalized,faulkner2017locally}.
These Laplace-like prior distributions can be expressed as scale mixtures of normal distributions with independent
gamma distributed variances \citep{kyung2010penalized}; for a more comprehensive discussion on shrinkage priors of this form, we refer readers to \citep{faulkner2017locally}.

The posterior $\text{P}(\boldsymbol{\alpha} | \mathbf{Y})$ can be presented in the following manner:
\begin{align*}
\text{P}(\boldsymbol{\alpha} | \mathbf{Y}) &\propto \text{P}(\mathbf{Y} | \boldsymbol{\alpha}) \text{P}(\boldsymbol{\alpha})\\
&\propto \prod_{i=1}^{500} \prod_{j=1}^{149} \text{P}(\mathbf{Y}_{i,j,\bullet} | \boldsymbol{\alpha}_{j,\bullet}) \prod_{l=1}^p \text{P}(\boldsymbol{\alpha}_{\bullet,l}).
\end{align*}
Note that we assume the $\mathbf{Y}_{i,j,\bullet}$ vectors are independent conditional on $\boldsymbol{\alpha}_{j,\bullet}$ and the prior distribution on $\boldsymbol{\alpha}$ factorizes across $l \in \{ 1, ..., p \}$.
The MAP estimate of the ensemble weights $\boldsymbol{\alpha}$ is obtained by maximizing $\text{P}(\boldsymbol{\alpha} | \mathbf{Y})$ and is equivalent to the estimate that minimizes the regularized objective function shown in the methods section.
The latter assertion can be seen as the posterior on $\boldsymbol{\alpha}$ can be monotonically transformed into our penalized minimization problem (up to a constant factor in $\boldsymbol{\alpha}$).

Of course, there are limitations to this interpretation of our modeling framework.
For instance, $\mathbf{Y}_{i,j,\bullet}$ is a frequency vector, yet we model the likelihood of $\mathbf{Y}_{i,j,\bullet}$ using a multivariate normal distribution, which has support over all real numbers; we could potentially remedy this problem by modeling the likelihood of $\mathbf{Y}_{i,j,\bullet}$ as a Dirichlet distribution as its negative log-likelihood looks similar to a cross-entropy loss function.
In addition, we specify priors on $\boldsymbol{\alpha}$ that also have support over the real line, which is not realistic.
Our assumption that the $\mathbf{Y}_{i,j,\bullet}$ vectors are conditionally independent given $\boldsymbol{\alpha}_{j,\bullet}$ is used solely for presentation purposes and does not hold in practice because the substitution profile data in both $\mathbf{X}$ and $\mathbf{X}^*$ are correlated across AHo positions.
Despite these issues with our statistical representation of the penalized regression model, our results demonstrate that the model is useful for predicting CF-specific substitution profiles in data-sparse situations.

\subsection*{Smoothed Jaccard Similarity}
As we stated in the methods section, optimization on the Jaccard similarity objective function is difficult because this metric is locally flat with respect to our parameter values of $\boldsymbol{\alpha}$.
For some small changes in $\boldsymbol{\alpha}$, the averaged Jaccard similarity can remain at the same value because the Jaccard sets continue to hold the same elements.
This is a problem because the L-BFGS-B optimization algorithm uses gradient information to determine its search direction for $\boldsymbol{\alpha}$ and the Jaccard similarity gradients are often zero due to the reasoning given above, which results in premature termination of the L-BFGS-B optimizer.
We now describe an approach to ``smooth'' the Jaccard similarity objective function that directly addresses these concerns.

\begin{figure}[ht!]
\centering
\includegraphics[width=0.6\textwidth]{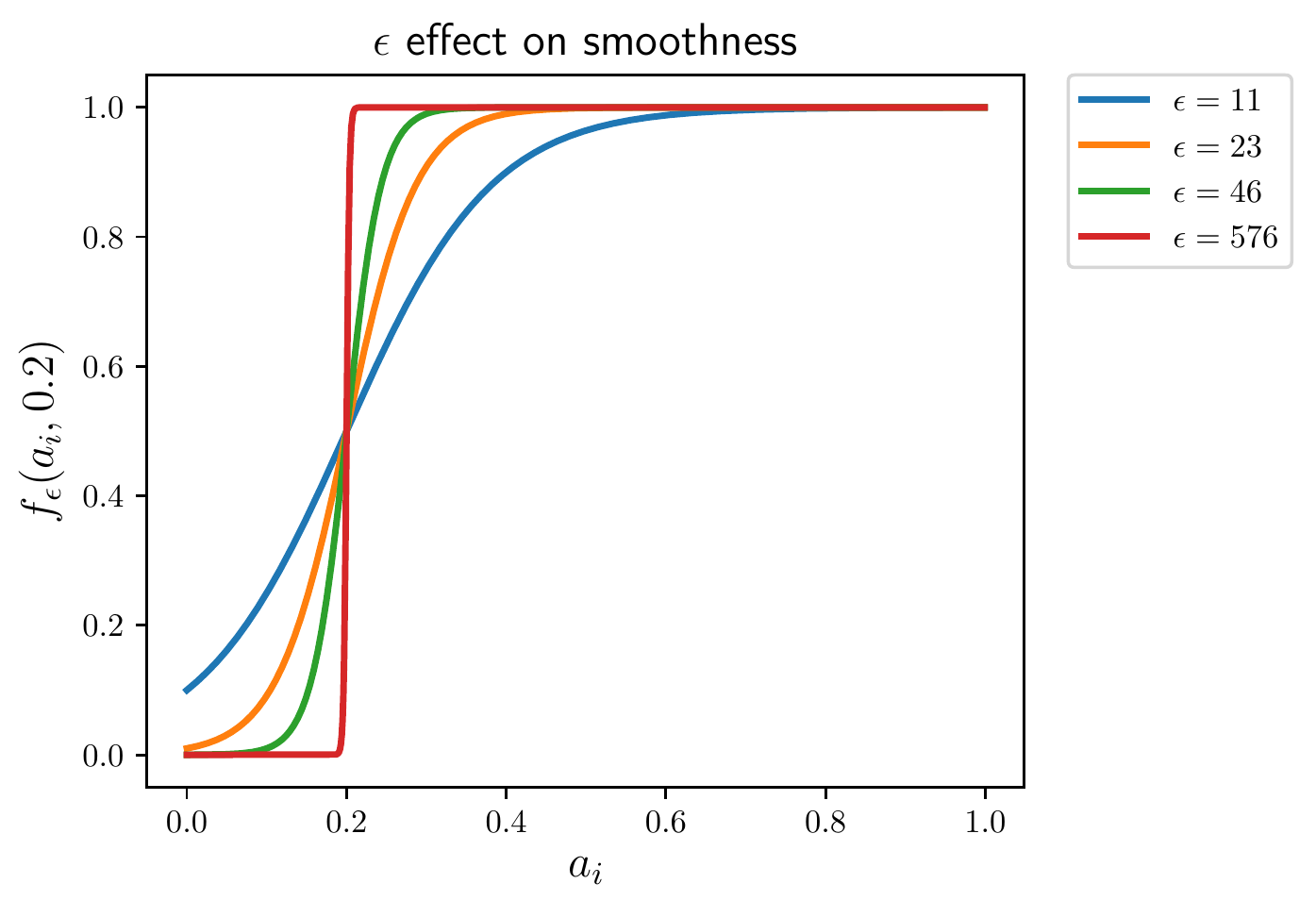}
\caption{A plot of the function $f_{\epsilon}(a_i, 0.2)$ against $a_i \in [0,1]$ for various values of $\epsilon$.
As $\epsilon$ gets larger, $f_{\epsilon}(a_i, 0.2)$ tends to the indicator function $f(a_i, 0.2)$.
}
\label{fig:smooth_jacc}
\end{figure}

For notational simplicity, we let $\{ a_i \}_{i=1:20}$ and $\{ b_i \}_{i=1:20}$ denote the actual and predicted amino acid frequencies, respectively, at a particular AHo position for a given CF.
As before, $t$ represents the cutoff separating high versus low frequency amino acids.
We also introduce the following indicator function $f(a, t) \equiv \mathbbm{1}\{ a \geq t \}$ for any amino acid frequency $a$.
If we further let $A = \mathcal{A}(\{a_i \mid a_i \geq t\})$ and $B = \mathcal{A}(\{b_i \mid b_i \geq t\})$ with $\mathcal{A}(\cdot)$ as defined in the methods section, then the Jaccard similarity between sets $A$ and $B$ can be rewritten as:
\begin{equation*}
J(A,B) \equiv \frac{|A \cap B|}{|A \cup B|} = \frac{\sum_{i=1}^{20} f(a_i, t) f(b_i, t)}{\sum_{i=1}^{20} \min\bigl\{1, f(a_i, t) + f(b_i, t)\bigr\}}.
\end{equation*}
The local flatness of the Jaccard similarity objective is due to the constant regions of $f(a_i, t)$ and the non-smooth curvature of $f(a_i, t)$ at the jump point $t$.
It turns out that $f(a_i, t)$ can also be described as the limit of the following function:
\begin{equation*}
f_{\epsilon}(a_i, t) = \frac{1}{1 + e^{-\epsilon(a_i - t)}},
\end{equation*}
as $\epsilon \rightarrow \infty$.
Thus, to obtain a ``smooth'' transformation of $J(A,B)$, we replace $f(a_i, t)$ with $f_{\epsilon}(a_i, t)$ in the above equation of $J(A,B)$ and set $\epsilon$ (i.e.\ the steepness parameter) to be a small number.
\autoref*{fig:smooth_jacc} plots the function $f_{\epsilon}(a_i, 0.2)$ against $a_i \in [0,1]$ for various values of $\epsilon$.

Fortunately, the use of this ``smooth'' Jaccard similarity function allows the L-BFGS-B optimization algorithm to converge properly.
To use this ``smoothed'' objective function in the right manner, we were interested in finding the largest values of $\epsilon$ that still permitted proper L-BFGS-B convergence.
We utilized $\epsilon = 23$ throughout all our Jaccard similarity experiments because we found that this value of $\epsilon$ satisfied our selection criteria specified previously.

\subsection*{Supplementary Figures/Tables}

\begin{table}[!ht]
\centering
\begin{tabular}{lccccc}
 & & & & & \\[-9pt]
Objective Function & $\widehat{\mathbf{X}}^*$ & $\widehat{\lambda}_1$ & $\widehat{\lambda}_2$ & $\widehat{d}$ & \text{Regularized CV} \\
\hline
 & & & & & \\[-9pt]
$L_2$ Error & $\bigl\{\widehat{\mathbf{X}}_{\naiveAA}, \widehat{\mathbf{X}}_{\vgene}, \widehat{\mathbf{X}}_{\neut}, \widehat{\mathbf{X}}_{\vsubgrp}\bigr\}$ & $10^{-7}$ & $10^{-5}$ & $3$ & 0.0453 \\
\hline
 & & & & & \\[-9pt]
Jaccard Similarity & \multirow{2}{*}{$\bigl\{\widehat{\mathbf{X}}_{\naiveAA}\bigr\}$} & \multirow{2}{*}{$10^{-7}$} & \multirow{2}{*}{$10^{-7}$} & \multirow{2}{*}{$2$} & \multirow{2}{*}{0.9316}\\
($t = 0.2$) &
\end{tabular}
\caption{The results from fitting the regularized models using 5-fold cross-validation.
We present the optimal tuning parameters selected from $\lambda_1, \lambda_2 = 10^{-7}, 5.05 \times 10^{-6}, 10^{-5}$ and $d = 1, 2, 3$ and show the associated cross-validated performance estimates.
Note that the possible choices of $\mathbf{X}^*$ for the $L_2$ error metric include the $\bigl\{\widehat{\mathbf{X}}_{\naiveAA}, \widehat{\mathbf{X}}_{\vgene}, \widehat{\mathbf{X}}_{\neut}\bigr\}$ and $\bigl\{\widehat{\mathbf{X}}_{\naiveAA}, \widehat{\mathbf{X}}_{\vgene}, \widehat{\mathbf{X}}_{\neut}, \widehat{\mathbf{X}}_{\vsubgrp}\bigr\}$ groupings, while the $\bigl\{\widehat{\mathbf{X}}_{\naiveAA}\bigr\}$ and $\bigl\{\widehat{\mathbf{X}}_{\naiveAA}, \widehat{\mathbf{X}}_{\vgene}\bigr\}$ groupings are the possible $\mathbf{X}^*$ choices for the smoothed Jaccard similarity objective.
}
\label{table:cv}
\end{table}

\end{document}